\documentclass[aps,twocolumn,prx,showpacs,psfig,superscriptaddress,longbibliography]{revtex4-2}
\usepackage{amsmath,amssymb}
\usepackage{braket}
\usepackage{bm}
\usepackage{color}
\usepackage{etoolbox}
\usepackage{epstopdf}
\usepackage{float}
\usepackage{graphicx}
\usepackage[colorlinks=true,linkcolor=blue,citecolor=blue,urlcolor=blue]{hyperref}
\usepackage[latin9]{inputenc}
\usepackage{lipsum}
\usepackage{multirow}
\usepackage{soul}
\usepackage{subfigure}
\usepackage{times}
\usepackage{titlesec}
\usepackage{tikz}
\usepackage{txfonts}
\usepackage{xspace}
\usepackage{xfrac}
\usepackage{ulem}
\usepackage[title]{appendix}

  \newcommand{\Sec}[1]{Sec.~\ref{#1}}
  \newcommand{\App}[1]{Appendix~\ref{#1}}
  \newcommand{\Eq}[1]{Eq.~\eqref{#1}}

  \newcommand{\Fig}[1]{Fig.~\ref{#1}}

  \def\Dstar{\ensuremath{D^\ast}\xspace}

  \def\iu{\ensuremath{\mathrm{i}\mkern1mu}}
  \def\Ul{\ensuremath{U_{c1}}\xspace} 
  \def\Uh{\ensuremath{U_{c2}}\xspace}

  \def\Cz{\ensuremath{{Q}}\xspace} 

\graphicspath{{./Figs/}}

\begin{document}
\title{Quantum Spin Liquid with Emergent Chiral Order
in the Triangular-lattice Hubbard Model}

\author{Bin-Bin Chen}
\affiliation{School of Physics, Beihang University, Beijing 100191, China}
\affiliation{Arnold Sommerfeld Center for Theoretical Physics, Center for NanoScience, 
and Munich Center for Quantum Science and Technology, Ludwig-Maximilians-Universit\"at M\"unchen, 
80333 Munich, Germany}
\affiliation{Department of Physics and HKU-UCAS Joint Institute of Theoretical and Computational Physics, 
The University of Hong Kong, Pokfulam Road, Hong Kong SAR, China}

\author{Ziyu Chen}
\affiliation{School of Physics, Beihang University, Beijing 100191, China}

\author{Shou-Shu Gong}
\email{shoushu.gong@buaa.edu.cn}
\affiliation{School of Physics, Beihang University, Beijing 100191, China}

\author{D. N. Sheng}
\affiliation{Department of Physics and Astronomy, California State University, Northridge, California 91330, USA}

\author{Wei Li}
\email{w.li@itp.ac.cn}
\affiliation{CAS Key Laboratory of Theoretical Physics, Institute of Theoretical Physics,
Chinese Academy of Sciences, Beijing 100190, China}
\affiliation{School of Physics, Beihang University, Beijing 100191, China}

\author{Andreas Weichselbaum}
\email{weichselbaum@bnl.gov}
\affiliation{Department of Condensed Matter Physics and Materials
Science, Brookhaven National Laboratory, Upton, New York 11973-5000, USA}
\affiliation{Arnold Sommerfeld Center for Theoretical Physics, Center for NanoScience, 
and Munich Center for Quantum Science and Technology, Ludwig-Maximilians-Universit\"at M\"unchen, 
80333 Munich, Germany}

\begin{abstract}
The interplay between spin frustration and charge fluctuation gives rise to an
exotic quantum state in the intermediate-interaction regime of the 
half-filled triangular-lattice Hubbard model, while the nature of the 
state is under debate. Using the density matrix renormalization group
with SU(2)$_{\rm{spin}} \otimes $U(1)$_{\rm{charge}}$ 
symmetries implemented, we study the triangular-lattice Hubbard model defined on 
the long cylinder geometry {up to circumference $W=6$}. 
A gapped quantum spin liquid, with on-site interaction $9 \lesssim U / t \lesssim 10.75$, 
is identified between the metallic and the antiferromagnetic Mott insulating phases.
In particular, we find that this spin liquid develops a robust long-range spin 
scalar-chiral correlation as the system length $L$ increases, 
which unambiguously unveils the spontaneous time-reversal symmetry 
breaking. In addition, the degeneracy of the entanglement spectrum
supports symmetry fractionalization and spinon edge modes 
in the obtained ground state. The possible origin of chiral order in this 
intermediate spin liquid and its relation to the rotonlike excitations 
have also been discussed.
\end{abstract}

\date{\today}
\maketitle

\section{Introduction}

Since Anderson's seminal work
of the resonating valence bond {(RVB)} state  in quantum 
antiferromagnets~\cite{Anderson1973,Anderson1987}, 
searching for spin liquid states and the consequent superconductivity after doping, 
constitutes an exciting topic in condensed matter physics~\cite{wen2006}.
While it has been widely accepted that spin frustration plays the key role for the 
emergence of spin liquid in Mott insulators~\cite{balents2010, savary2016, zhou2017},
it has also been noticed that the strong charge 
fluctuations near the Mott transition may add an additional 
active ingredient to the system~\cite{Motrunich2005,lee2005}.
Although stable spin liquid states in the half-filled 
bipartite-lattice Hubbard models have not been established~\cite{Sorella2012}, 
the frustrated triangular-lattice Hubbard (TLU) model, 
harboring stronger spin and charge fluctuations
{at intermediate Hubbard interaction $U$},
has raised great interests in the possible intermediate spin liquid state
\cite{Morita2002,Koretsune2007,Sahebsara2008,Yoshioka2009,Shirakawa17,
Szasz20,Szasz21}.

Meanwhile, experimental progress in the triangular-lattice 
organic-salt compounds $\kappa$-(BEDT-TTF)$_2$Cu$_2$(CN)$_3$ 
\cite{Shimizu2003,Kurosaki2005,Yamashita2008,Isono2014,Isono16,Miksch21} 
and EtMe$_3$Sb{$[$}Pd(dmit)$_2${$]$}$_2$ \cite{Yamashita2010,Yamashita2011}
also shed light on the spin-liquid states near the Mott transition.
The absence of spin ordering down to the 
lowest experimental temperature and the linear-$T$ 
dependence of low-temperature specific heat 
suggest a possible gapless spin liquid in these compounds~\cite{Yamashita2008,Yamashita2011}.
However, recent thermal conductivity measurements indicate 
the absence of mobile gapless excitations~\cite{hope2019, ni2019}. 
The experimental identification of the spin liquid and the pursuit
of its nature have further stimulated intensive theoretical studies.

To include the charge fluctuation effects, one can consider the higher-order 
ring-exchange coupling in the effective spin model
~\cite{Motrunich2005,Schroeter07,Sheng2009,Yang2010,Cookmeyer21}
or simulate the Hubbard model directly. 
Indeed, numerical simulations on the ring-exchange spin model have 
identified a gapless spin liquid state with the emergent 
spinon Fermi surface~\cite{Motrunich2005,Sheng2009,Block2011}, 
which can partly explain the experimental findings. 
On the other hand, large-scale density matrix renormalization
group (DMRG) simulations on the TLU itself
have uncovered a spin liquid phase near the Mott transition
\cite{Shirakawa17,Szasz20}.
However, the two different studies
lead to drastically distinct conclusions
on the nature of this spin liquid.
While the finite-DMRG calculation \cite{Shirakawa17} suggests 
a Dirac-like gapless spin liquid preserving time reversal symmetry (TRS), 
the more extensive infinite-DMRG study \cite{Szasz20}
finds a gapped chiral spin liquid (CSL)
with finite chiral order~\cite{Kalmeyer87,Bauer14,Hu2015,Gong2019}.
Moreover, the spinon Fermi-surface state is not found 
in these DMRG simulations, in contrast to the previous understanding 
based on the effective spin model.

\begin{figure}[htb]
\includegraphics[width=1\linewidth]{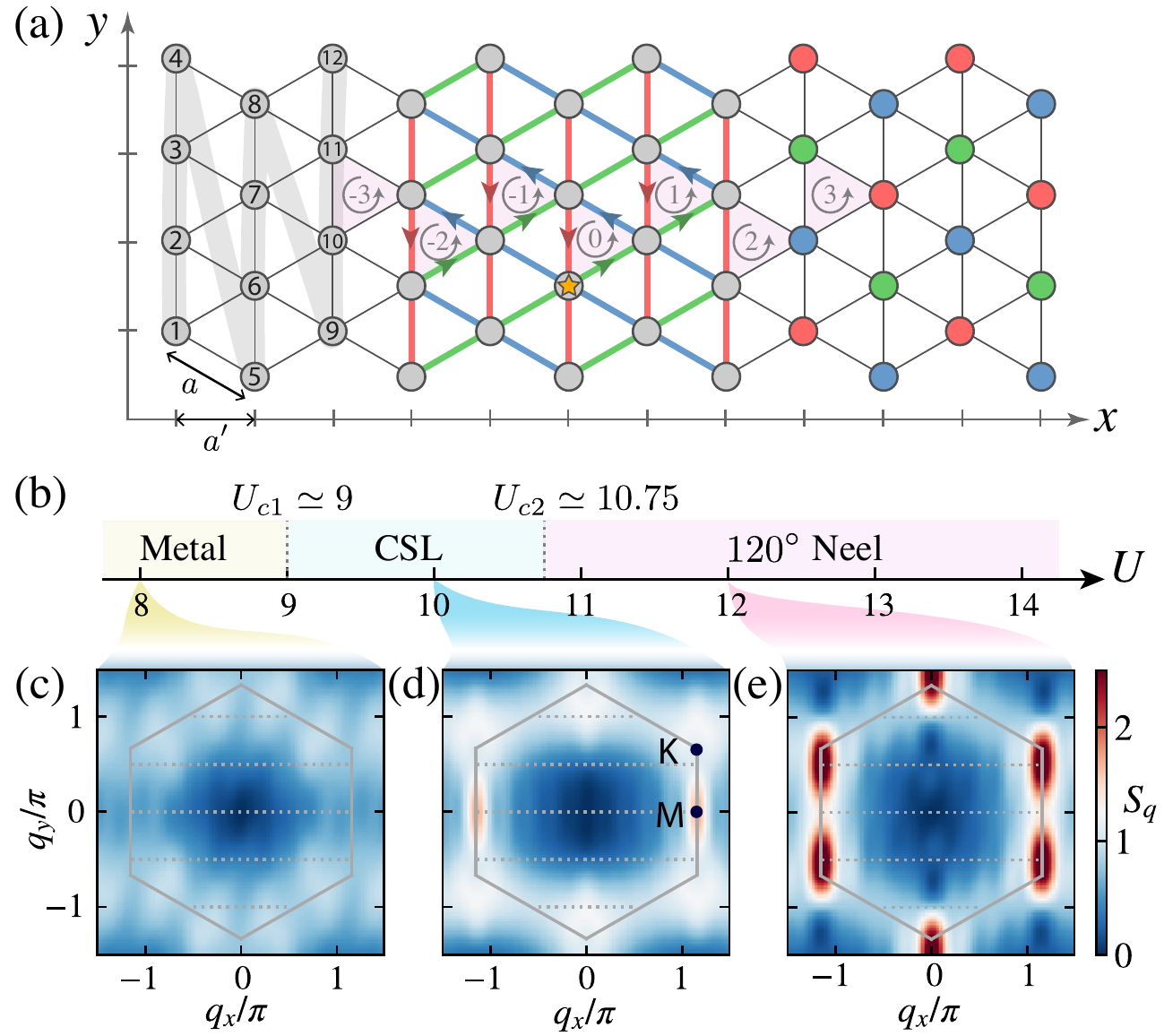}
\caption{Model and phase diagram. 
(a) Triangular-lattice cylinder with open (periodic) 
boundary conditions along the $x$ ($y$) direction
and lattice spacing $a=1$. 
{Having a straight open edge, i.e., YC4 geometry,
the columnar spacing is $a'=\sqrt{3}/2$.
{(For later reference, an XC geometry has 
straight horizontal chains at distance $a'$
with a zigzag open boundary left and right).}}
The DMRG simulations employ a 1D ``zigzag" 
mapping as indicated by the gray-shaded path 
(with part of the site ordering also indicated). 
The yellow star denotes the central reference site 
for which the spin-spin correlation is calculated, 
while the pink triangles for evaluating the chiral correlations 
are ordered symmetrically away from the center.
The arrows on the bonds denote the current directions
of the chiral order, having three colors for the
three different directions.
(b) The phase diagram of the TLU model consists of a metallic phase, 
a fully gapped CSL phase, and a 120$^\circ$ spin-ordered phase,
with the three colors of the sites denoting
the three-sublattice structure. 
(c)-(e) Typical static spin structure factors $S_q$ for $U=8, 10$, and 12
in the three phases.
}
\label{Fig:Phase}
\end{figure}

In this work,
we further determine the precise nature of the spin liquid
phase in the TLU. After introducing the model and method in 
\Sec{sec:model}, we perform extensive DMRG calculations 
on finite-size cylinders for a fixed width $W = 4$
(YC4; cf. \Fig{Fig:Phase}) in \Sec{sec:YC4}, 
where we gradually increase the system length up to 
$L = 64$. This goes far beyond the previous finite-size
DMRG \cite{Shirakawa17} and thus significantly reduces 
finite-size effects, {which is followed by an analysis of
width $W=6$ cylinders (YC6) in \Sec{sec:YC6}, with strong
evidence also for a chiral phase there, albeit with slightly
altered phase boundaries.}
Throughout, we emphasize the necessity to
exploit the SU(2)$_{\rm{spin}} \otimes$ U(1)$_{\rm{charge}}$ 
symmetries in our DMRG simulation, as this 
permits us to reliably reach large-scale systems.
We identify an intermediate non-magnetic phase with
ultrashort single-particle and spin correlation lengths
on the order of one lattice spacing,
having $(\Ul \simeq 9 t) \lesssim U \lesssim
(\Uh \simeq 10.75 t)$ {for $W=4$}.
On short cylinders we find
exponentially decaying chiral correlation
in agreement with Ref.~\cite{Shirakawa17},
but the result changes fundamentally with increasing 
system length,
showing very robust long-range chiral correlation characterizing
spontaneous TRS breaking.
We also find a large degeneracy in the entanglement spectrum, 
which agrees with symmetry fractionalization and
the existence of an edge spinon in the obtained ground state. 
Therefore, we conclude that the low-energy physics of
the TLU model at the intermediate-$U$ is governed by a gapped CSL.
Our results show the importance 
of a sufficiently large system length to overcome 
finite-size effects for identifying TRS breaking in this system.
In the outlook {in \Sec{sec:summary}},
we point out a possible link between
the present results and the chiral nature of excitations
found in the {triangular lattice} Heisenberg {(TLH)}
model at finite temperature \cite{Chen2019}.

\section{Model and Method}
\label{sec:model}

The TLU model is defined as
\begin{equation}
\label{Eq:Hub}
\hat{\mathcal{H}} = -t\sum_{\langle i,j \rangle, \sigma} (\hat{c}_{i\sigma}^\dagger \hat{c}^{\phantom{\dagger}}_{j\sigma} + \mathrm{H.c.}) 
    + \tfrac{U}{2}(\hat{n}_i - 1)^2,
\end{equation}
where $\langle \cdot,\cdot \rangle$ represents the summation over 
the nearest-neighbor (NN) couplings, $\hat{c}^{\,}_{i \sigma}$ 
($\hat{c}^{\dagger}_{i \sigma}$) denotes the fermionic annihilation 
(creation) operator with spin $\sigma\in\{\uparrow,\downarrow\}$ 
on site $i$, and $\hat{n}_i = \sum_\sigma \hat{n}_{i \sigma}$ with
$\hat{n}_{i\sigma} {\equiv} \hat{c}_{i\sigma}^\dagger \hat{c}^{\phantom{\dagger}}_{i\sigma}$ 
the particle number operator. We set $t:=1$
as the unit of energy, {and the unit of distance
via the lattice spacing $a:=1$, throughout}.
In our DMRG calculations \cite{White1992},
we mainly focus on YC4 cylinders as shown in \Fig{Fig:Phase}(a),
but also extend to {XC4 and} YC6 cylinders. Throughout, we
implement the $\mathrm{SU}(2)_{\mathrm{spin}} \otimes
\mathrm{U}(1)_{\mathrm{charge}}$ 
symmetries based on 
the QSpace tensor library~\cite{Weichselbaum2012,Weichselbaum2020}
which enables us to retain up to $D^* = 8\,192$ multiplets 
[equivalent to about ${D\sim}24\,000$ U(1) states] and 
ensures full convergence with a 
truncation error $\lesssim$
$1\times 10^{-6}$.

\begin{figure}[tb]
\includegraphics[width=.99\linewidth]{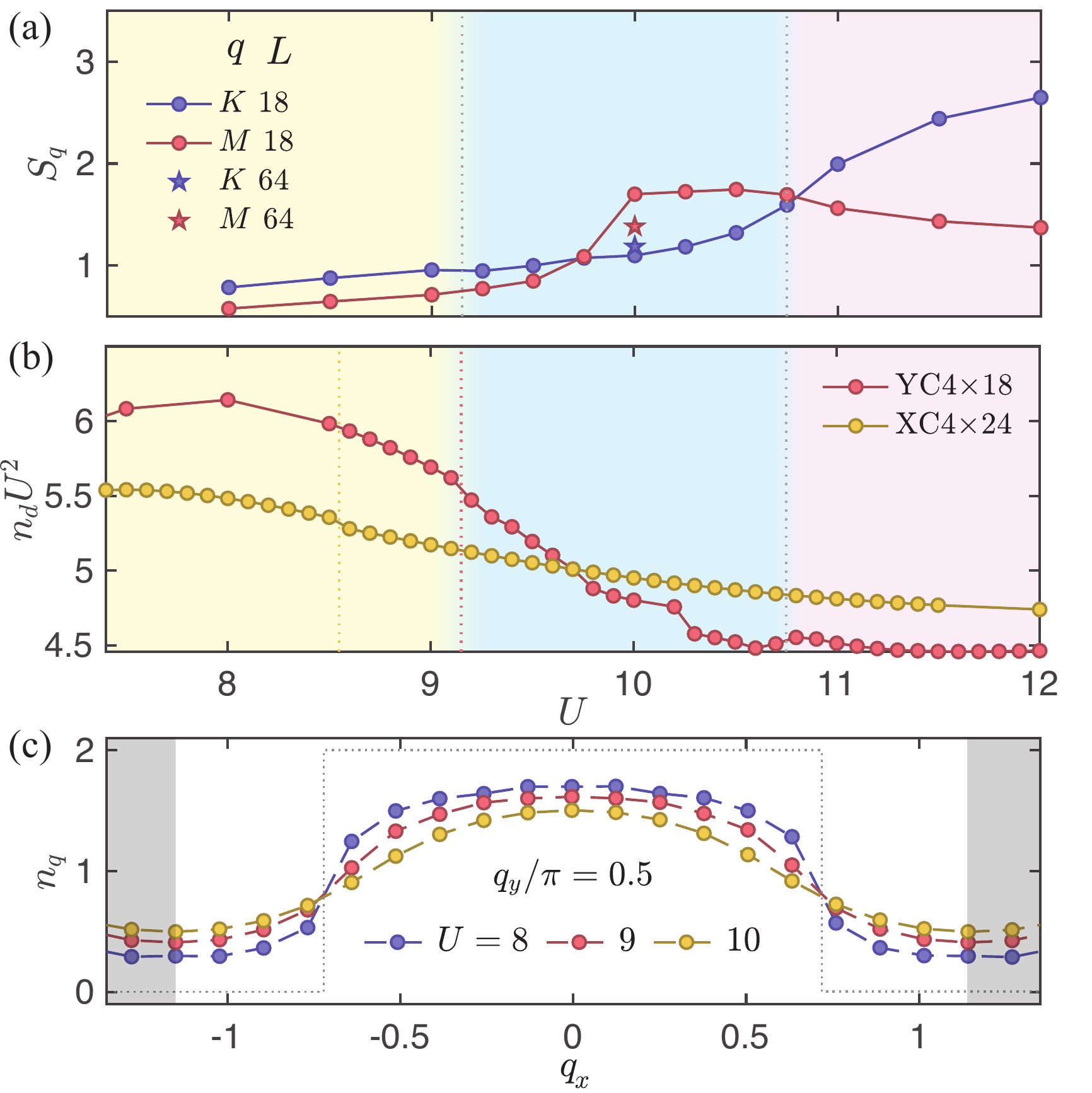}
\caption{Hubbard $U$ dependence of different quantities 
on the YC4$\times${$L$} cylinders. 
(a) Spin structure factors $S_q$ at the $K$ and $M$ 
points. The $L=64$ data are also shown at $U=10$.
{The asterisks show data points for the significantly
longer system that exhibits long-range chiral correlations
as analyzed in \Fig{Fig:Chiral}.}
(b) {Double occupancy shown as} $n_d U^2$.
{This also includes XC4 data which exhibits 
only one discontinuity around the metal-insulator transition.} 
(c) Distribution of $n_q$ vs. $q_x$
in the momentum space along $q_y = \pi / 2$.
The dashed line plots $n_q$ for $U = 0$ on the torus 
system with the same size $4\times 18$.
The white center region marks the first Brillouin zone.
}
\label{Fig:Obs}
\end{figure}

\section{Width $4$ Cylinders (YC4)}
\label{sec:YC4}

\subsection{Ground-state phase diagram} 

We summarize our DMRG 
phase diagram in \Fig{Fig:Phase}(b), 
with a metallic phase for $U < \Ul ({\simeq} 9)$, 
a $120^{\circ}$ magnetically coplanar-ordered phase
\cite{Henley89,White2007,Wb11} for $U > \Uh ({\simeq} 10.75)$, 
and an intermediate CSL phase with spontaneous TRS breaking. 
\Fig{Fig:Phase}(c-e) show {the} representative
snapshots of the static
spin structure factors 
$S_q \equiv \sum_j 
e^{\iu q\cdot R_{0j}}
\langle \hat{S}_{j} \cdot \hat{S}_{0} \rangle$
in the different phases
where site $0$ refers to a fixed site in the 
center of the system [c.f., the 
asterisk site in Fig.~\ref{Fig:Phase}(a)].
In the metallic phase (small $U$) $S_q$ is
found to be featureless. In the CSL phase [\Fig{Fig:Phase}(d)]
a peak emerges at the $M$ point
which is related to short-range stripe correlation.
Further increasing $U$, sharp peaks emerge around 
the $K$ points [\Fig{Fig:Phase}(e)], 
consistent with a semiclassical 120$^\circ$ spin order.
Even though the present YC4 geometry is not fully compatible with the 120$^\circ$ order, 
the feature of the dominant $K$ point peak can still be observed.
For very large $U\gtrsim 20$, eventually,
the ground state on the particular YC4 cylinder 
switches to an RVB ring-like
state \cite{Chen2018}, which is beyond the scope of interest here.

The phase boundaries of the CSL phase are estimated
in \Fig{Fig:Obs}. 
By contrasting the $U$ dependence of $S_q$ at $q=K$ with 
$M$ in \Fig{Fig:Obs}(a)
\footnote{although the Brillouin zone of the YC4 system
does not contain the $K$ point, 
we can still compute the Fourier transform of spin 
correlations at this momentum, as a proper approximation},
strongly enhanced magnetic correlations
at the $M$ points appear in the 
intermediate regime,
up to $L=64$.
For $L=18$, this leads
to two crossing points of $S_M$ with
$S_K$. We use the upper crossing
to estimate the phase boundary towards
the 120$^\circ$ order, resulting in $\Uh \simeq10.75$.
The lower phase boundary towards the metallic phase
represents a metal-insulator transition, 
that is more naturally characterized
by an analysis of the double
occupancy $n_d \equiv \langle\psi| 
\hat{n}_{0\uparrow} \hat{n}_{0\downarrow} |\psi\rangle$,
{which is related to local charge or energy
fluctuations of a single site
(computed at the central site $i=0$ here).
Starting from half-filling, this expectation
value acquires a finite value for $U \gg t$
via a second order process.
In the infinite-$U$ limit, one can thus estimate
for the magnetically completely uncorrelated case
$n_d \simeq \tfrac{z t^2}{2 U^2}$,
with $z$ the coordination number. Antiferromagnetic
correlations tend to increase this value.
Also as $U$ is lowered, charge fluctuations generally
increase via higher-order processes.
Hence in the present case with $z=6$ and $t=1$,
we can consider $n_d U^2 \gtrsim 3$ a lower bound. 
Also from a numerical perspective with $U \sim 10$,
we have for the double occupancy $n_d \ll 1$,
such that accurate numerical simulations are important.}
Now when plotting $n_d U^2$ vs. $U$ {for the YC4 system}
[\Fig{Fig:Obs}(b)], we observe the onset of 
a kink around $U\simeq 9$, which we thus
interpret as the lower phase boundary \Ul. 
Note, however, that this value for \Ul
does not coincide with the lower crossing of $S_M$ vs. $S_K$ in \Fig{Fig:Obs}(a),
which may be due to the finite-size effects,
considering that the lower phase boundary is
significantly more demanding numerically
given that both, spin and charge gaps close there. 
{{For YC4}, the double occupancy
$n_d U^2$ shows another discontinuity
around the magnetic transition \Uh,
{and it stays rather constant thereafter.
For comparison, we also include data for the
XC4 geometry in \Fig{Fig:Obs}(b).} 
In contrast to YC4, 
this only exhibits one discontinuity around \Ul.
{Overall, it is smoother, varies
significantly less vs. $U$, and remains higher
for the largest $U$. Together with the fact
that 120$^\circ$ order is commensurate with XC4,
we take the above finding as an indication that
the intermediate chiral phase is absent for the
XC4 geometry. As such, it exhibits a lesser degree of
frustration and favors the large-$U$ magnetic
correlations already at smaller $U$. As an aside,
we note that the interpretation
that frustration is less pronounced for XC4 is also
supported by Ref. \cite{Wb11}, where no dimerization
was observed in the Heisenberg limit.
}}

The metal-insulator transition also manifests
itself in the change of the Fermi surface with increasing $U$.
For this, we analyze 
the one-particle charge density in 
momentum space $n_q 
\equiv \sum_{\sigma} \sum_j 
e^{\iu q\cdot R_{0j}}  
\langle \hat{c}_{j\sigma}^\dagger \hat{c}_{0\sigma}^{\,} \rangle$
in \Fig{Fig:Obs}(c).
At $U = 0$, $n_q$ is a step-like function (dashed line). 
With increasing $U$, the drop of $n_q$ is 
gradually smoothened until the Fermi surface disappears
\cite{Szasz20}. 
To be specific, within the resolution for $L=18$,
$n_q$ at $U = 8$ still exhibits an 
appreciable ``jump'' at $q_x \simeq \pm  3 \pi / 4$ with fixed
$q_y = \pi / 2$. However, for $U \gtrsim 9$, 
$n_q$ changes smoothly
and the Fermi surface appears to be absent
(cf. \App{app:Fermi:surface}),
indicating a metal-insulator transition.
Our estimates of the phase boundaries
are roughly consistent with those reported in previous studies
\cite{Yoshioka2009,Yang2010,Laubach15,Shirakawa17,Szasz20}.
Minor quantitative deviation of the lower boundary \Ul 
is likely due to the different geometry and system size.

\begin{figure}[t!]
\includegraphics[width=.95\linewidth]{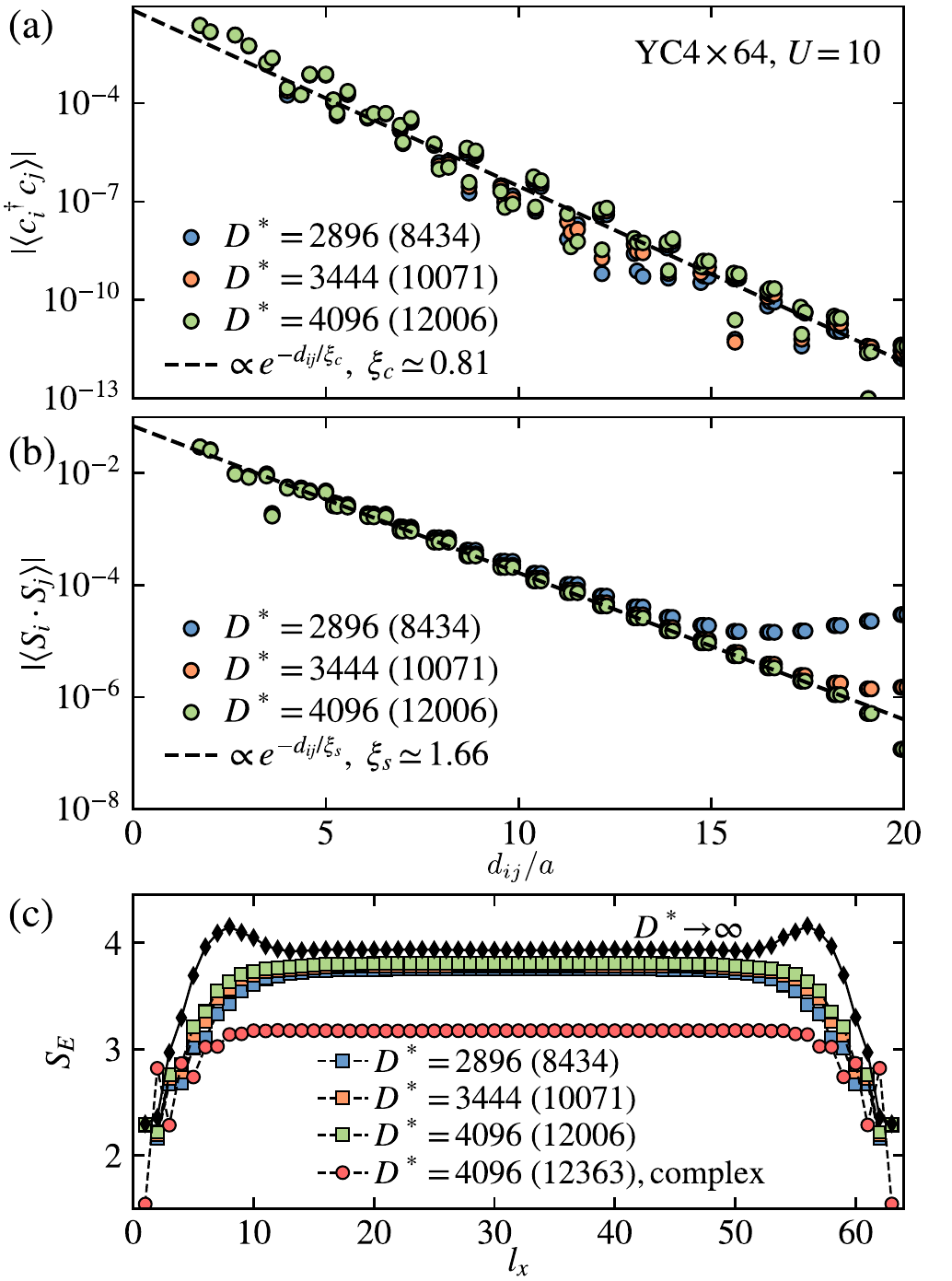}
\caption{
Single-particle Green's function, spin correlation, 
and entanglement entropy obtained on the YC4$\times$64 cylinder at $U = 10$. 
Log-linear plot of (a) single-particle Green's function $|\langle c^\dagger_i c^{\,}_j \rangle|$, 
and (b) spin correlation $|\langle \bm{S}_i \cdot \bm{S}_j \rangle|$ 
as a function of distance $d_{ij}$, which show exponential decay with short decay lengths 
${\xi_c} \simeq 0.81$ and ${\xi_s} \simeq1.66$, respectively.
(c) (c) Bipartite entanglement entropy $S_E = - \sum_i \rho_i \ln \rho_i$ 
for the reduced density matrix $\rho$ when cutting the system at bond $l_x$ 
vs. block size $l_x$.
This shows a well-developed plateau for $12 \lesssim l_x \lesssim 52$, 
and hence obeys the area law in the bulk.
We include a linear extrapolation $1/D^\ast {\to} 0$
(black symbols). $S_E$ is also obtained from a complex wavefunction
(red symbols; {all other data for a real wave function})
with the value of the plateau in the center
reduced by $\ln(2 {\pm 0.1})$. 
}
\label{Fig:Obs2}
\end{figure}

\subsection{Fully gapped spin liquid}

Next we focus on the charge and spin excitations in the 
spin liquid phase. 
In \Fig{Fig:Obs2}(a), the single-particle Green's function
$\langle \hat{c}^\dagger_i \hat{c}^{\,}_j \rangle$, averaged amongst all site pairs with 
the same distance $d_{ij}$, decays exponentially versus $d_{ij}$ with 
a short correlation length $\xi_c\simeq0.81$. 
This is consistent with a sizable charge (single-particle excitation) gap 
$\Delta_c \simeq 0.88$ extrapolated for $1/L\to0$
(see \App{app:finite-size}) 
and verifies that the spin liquid resides in the Mott insulator phase.

In \Fig{Fig:Obs2}(b), we show the spin correlation
$\langle \hat{S}_i \cdot \hat{S}_j \rangle$ versus $d_{ij}$,
which is well converged and clearly decays exponentially, 
with a short correlation length $\xi_s \simeq 1.66$. 
Such a short correlation length implies gapped spin excitations, 
which do not support a spinon Fermi surface state with 
algebraically decaying spin correlation~\cite{Motrunich2005} 
but could be consistent with either a gapped spin liquid 
or a Dirac-like gapless spin liquid that is gapped due to 
finite size on narrow-width cylinders.  As shown in~\Fig{Fig:Obs2}(c), 
this conclusion is further supported by the saturated bipartite 
entanglement entropy $S_{E}$ vs. subblock size $l_x$, following 
an area law~\cite{Calabrese2004a}.

{By analyzing the convergence of the DMRG simulation with
increasing $\Dstar$ (see Appendices \ref{app:convergence} and
\ref{app:finite-size} for details),
we find that the chiral
correlations become well-established only once the
accuracy of the energy per site reaches a resolution of
$\Delta e_\chi \cong e_g - e_g^0 \simeq 10^{-3}$
for YC$4\times 64$ [e.g., see \Fig{Fig:Conv}(h),
where the kink in the convergence of the ground state
energy around $1/\Dstar \sim 5\cdot10^{-4}$
with $\delta e_g \simeq 10^{-3}$ relates to a significant
build-up of entanglement entropy in \Fig{Fig:Conv}(g)
still, before it converges to a plateau].
Based on this, one may estimate
a sizable bulk gap
$\Delta_\chi \simeq N \cdot \Delta e_\chi \sim 0.26$
with the chiral phase,
with $N \equiv L\, W$ the total number of sites.
Nevertheless, since large system sizes are required
to capture the chiral phase,
the bulk gap's {\it relative} effect on the ground-state energy is
very small, thus requiring an energy accuracy
of at least $0.2\%$ for the case of
YC$4\times 64$.
Therefore, sufficiently accurate simulations on large
systems are important.
In the present case, this is made possible
by fully exploiting the SU(2) spin symmetry.
The sizable bulk gap $\Delta_\chi$ is consistent with
the extremely short spin and charge correlation
lengths on the order of the lattice spacings
itself, as seen in \Fig{Fig:Obs2}. 
Importantly, these correlation lengths are
already also much shorter than the width
of  the YC4 cylinder analyzed here.
In this sense, it appears plausible that
the chiral phase persists to wider systems in
the low-energy regime. And, indeed, as we will demonstrate
further below, we see
a consistent picture including chiral long-range
correlations also for the YC6 cylinder.}

\subsection{Spontaneous time-reversal symmetry breaking
\label{sec:TRS}} 

One key debate in the previous DMRG studies 
is whether there exists a spontaneous TRS 
breaking~\cite{Szasz20} or not~\cite{Shirakawa17}. 
Here we resolve this issue by calculating the spin 
chiral correlation 
$\langle \hat{\chi}_i \hat{\chi}_j \rangle$ 
{(see the Appendices for more details)}
between two three-spin triangles $\Delta_i$
and $\Delta_j$ symmetrically separated from
the system center, i.e., having $i$=$-j$ or 
$i$=1-$j$, as shown in \Fig{Fig:Phase}(a).
The involved scalar chirality operator
is $\hat{\chi}_i = (\hat{\sigma}_\alpha \times \hat{\sigma}_\beta) 
\cdot \hat{\sigma}_\gamma$
with $\alpha, \beta, \gamma \in \Delta_i$
in counter-clockwise order for the Pauli operators.
{Given that the Hermitian operator
$\hat{\chi}_i$ has purely imaginary matrix elements,
using real-valued DMRG with the real-valued Hamiltonian
in \Eq{Eq:Hub} will always yield $\langle
\hat{\chi}_i\rangle=0$. Evidently, a real-valued wave
function cannot (spontaneously) break TRS.
Hence we compute static chiral correlations
$\langle \hat{\chi}_i \hat{\chi}_j \rangle$.
Eventually, however, we do repeat precisely the same DMRG
calculations but using complex arithmetic, which then
permits a plain non-zero expectation value of
the order parameter $\langle \hat{\chi}_i\rangle$.}

\begin{figure}[tb]
\includegraphics[width=1\linewidth]{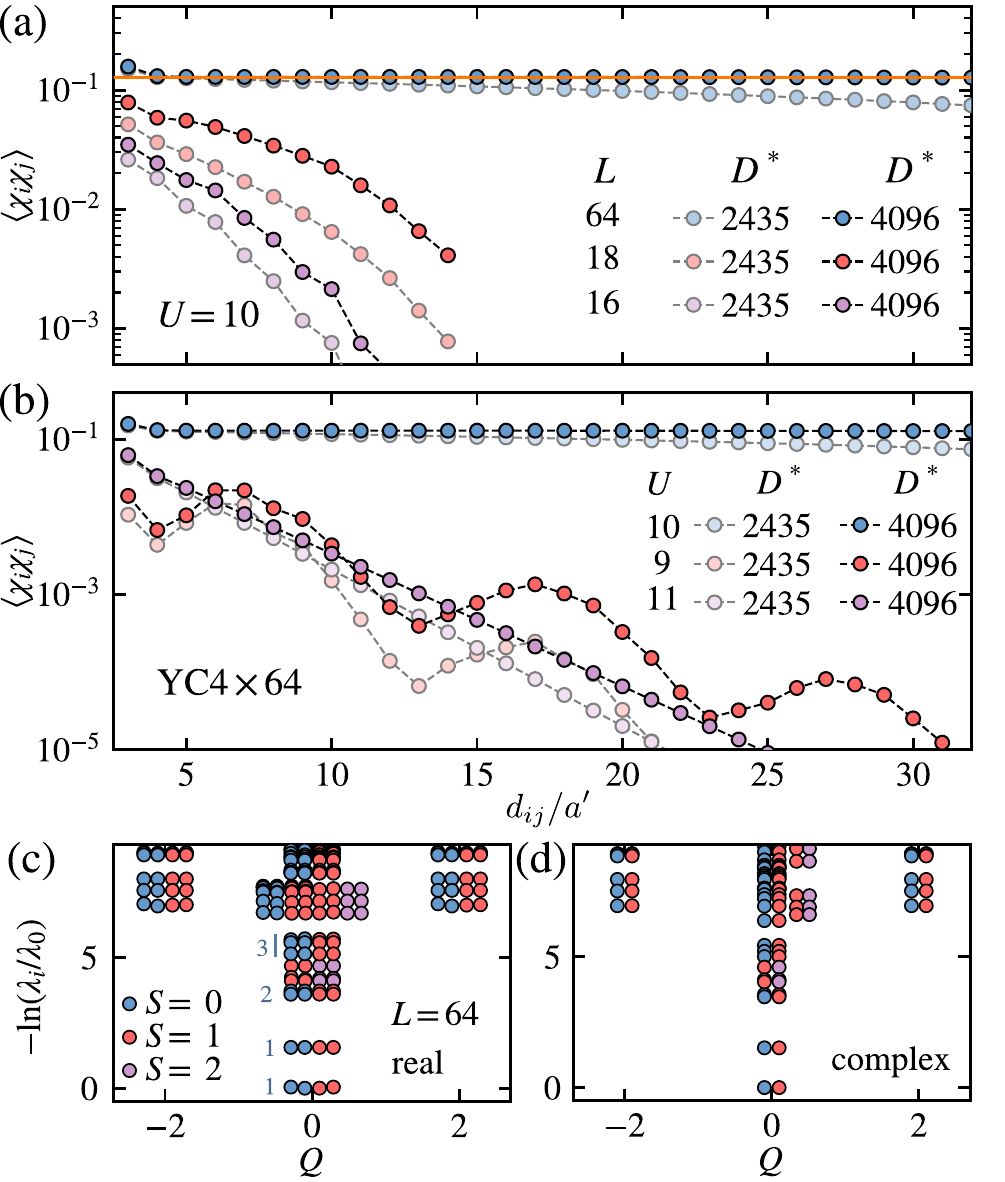}
\caption{
Chiral correlation and entanglement spectrum of the chiral spin liquid state.
(a) Chiral correlations $\langle \chi_i \chi_j \rangle$ for $U = 10$ 
on the YC$4$ cylinders with different system lengths $L = 16, 18, 64$. 
The orange horizontal line indicates the value of
$\langle\chi_i \chi_j \rangle{\simeq0.128}$ for $d_{ij}=30$.
(b) Chiral correlations for $U = 9, 10, 11$ on the YC$4\times64$ cylinder 
obtained by keeping the bond dimensions up to $D^*=4096$ multiplets.
Entanglement spectra of the YC$4\times64$ for both (c) real and 
(d) complex wavefunctions, grouped by charge sectors (for even \Cz),
with spin labels color-coded as specified in the legend.
The bars and respective numbers with the $Q=0$ column in (c)
indicate group degeneracy. 
The subtracted ground levels for the two cases are 
$\lambda_0\simeq0.08$ and $\lambda_0\simeq0.15$, 
where a relative factor of $\sim2$ is observed.
}
\label{Fig:Chiral}
\end{figure}

We first check the system-length dependence of the 
chiral correlation at $U=10$. 
As shown in \Fig{Fig:Chiral}(a), for small length $L = 16$ 
and $18$, the chiral correlations decay 
exponentially in the same sign, in agreement with
the previous study~\cite{Shirakawa17}.
Interestingly, by further increasing the system length,
a very robust chiral correlation is established 
over long distances,
with $\sqrt{\langle \hat{\chi}_i \hat{\chi}_j \rangle} \simeq 
0.36$ for $d_{ij}=30$ already well-converged
over distance for $D^\ast\gtrsim3444$ [c.f. the orange 
horizontal guide in \Fig{Fig:Chiral}(a)].
We also perform a complex-valued DMRG simulation 
{in the YC4$\times$64 system, to directly estimate}
the chiral order parameter 
$\langle \hat{\chi} \rangle\simeq0.35$ in the bulk, 
in excellent agreement with 
previous infinite-DMRG value
$\chi_\mathrm{iDMRG}(U{=}10)\simeq
0.34$~\cite{Szasz20}. Our results indicate 
that the system length is crucial for identifying 
the spontaneous TRS breaking in the DMRG 
calculations, which reconciles the different 
observations in previous studies. 
In \Fig{Fig:Chiral}(b) we also compute the
chiral correlations in the neighboring phases.
In either case, $(U{=}9) \lesssim 
\Ul$ and $(U{=}11) > \Uh$,
chiral correlations decay exponentially, consistent with the
preserved TRS in the two phases.

In addition, we also studied the XC cylinder, i.e., 
with one of the bond directions along the $x$-axis,
with circumference up to $W = 6$.
There, however, even for large $L$, we find no strong
signature of long-range chiral correlation
(see the {Appendices} for more details).
Such different behaviors of chiral correlation on 
different geometries have also been observed in the DMRG study of a triangular spin model 
with further-neighbor interactions~\cite{Gong2019},
where the chiral order on the XC cylinder 
emerges only at large circumference. We suspect
that long-range chiral correlation
ultimately also can be found on 
wider XC cylinder in future studies.

\begin{figure}[t!]
\!\!\includegraphics[width=1\linewidth]{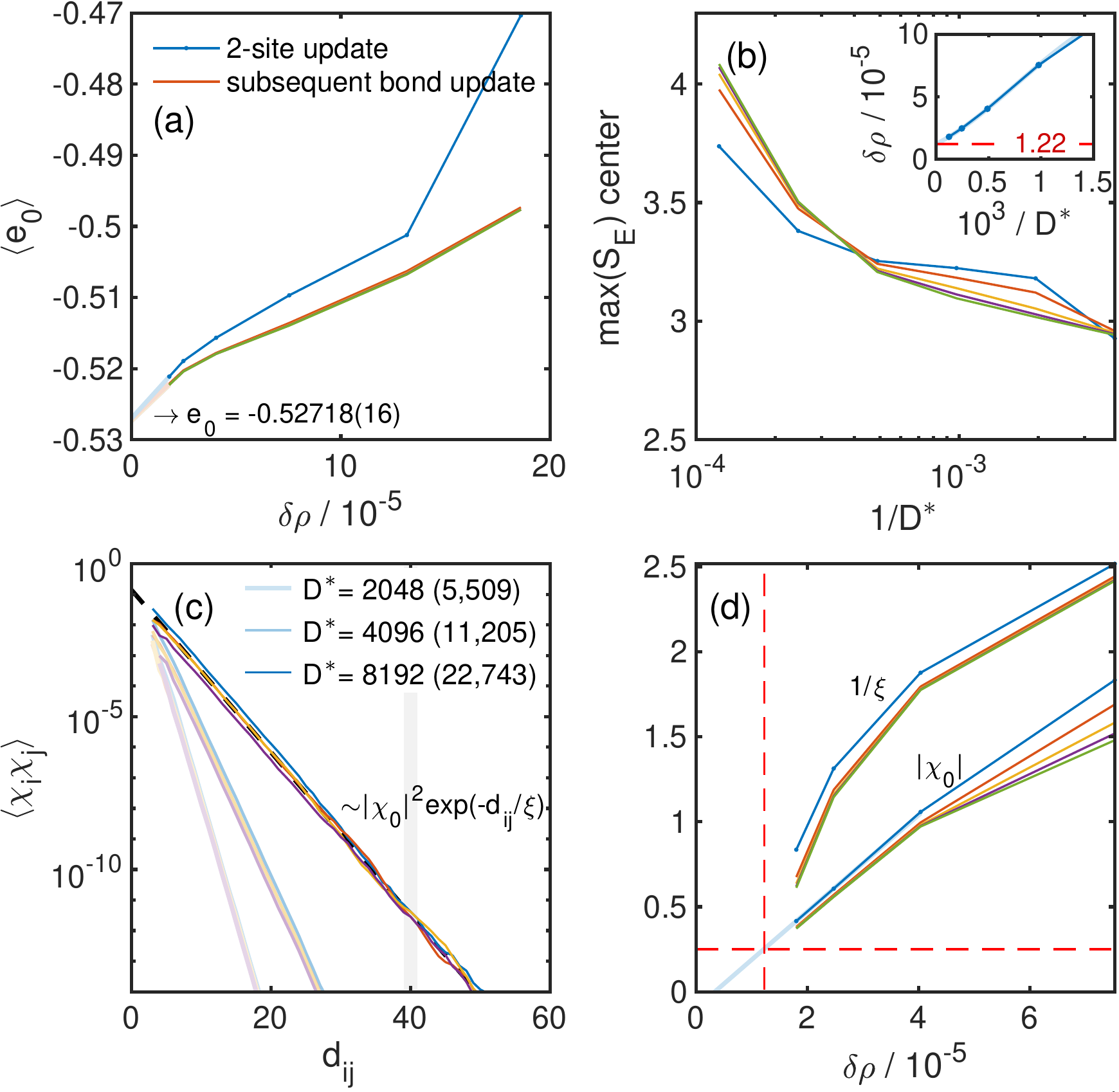}
\caption{{\ DMRG simulations (real-valued) on 
YC$6\times 64$\, at $U  = 9$. 
\mbox{(a) Convergence} of ground-state energy vs. 
    discarded weight $\delta\rho$. The lowest
    energies are extrapolated in a linear fashion
    towards $\delta\rho\to 0$, with the resulting 
    ground-state energy per site as shown.
(b) Maximum entanglement entropy around the system
    center vs. $1/\Dstar$, with $\Dstar$
    the number of multiplets
    kept in the simulation. The inset shows the
    discarded weight vs. $1/\Dstar$.
    The lines in panels (a) and (b) combine data from equivalent
    sweeps, such as the two-site update {(line with symbol)} when
    increasing \Dstar, or the subsequent four
    bond updates [for {all other lines,} see the legend to (a)].
(c) Chiral correlations vs. distance. While these data show 
    exponential decay, the correlation length is
    strongly dependent on $\Dstar$ still. Here the various
    lines are derived from different combinations of up
    and down triangles vs. distance. {Different
    levels of color intensity refer to different $\Dstar$ ($D$)
    as indicated in the legend.}
(d) Analysis of the parameters from the 
    exponential fits in (c) vs. discarded
    weight $\delta\rho$.
}}
\label{fig:YC6:real} 
\end{figure} 

\begin{figure}[t!]
\!\!\includegraphics[width=1\linewidth]{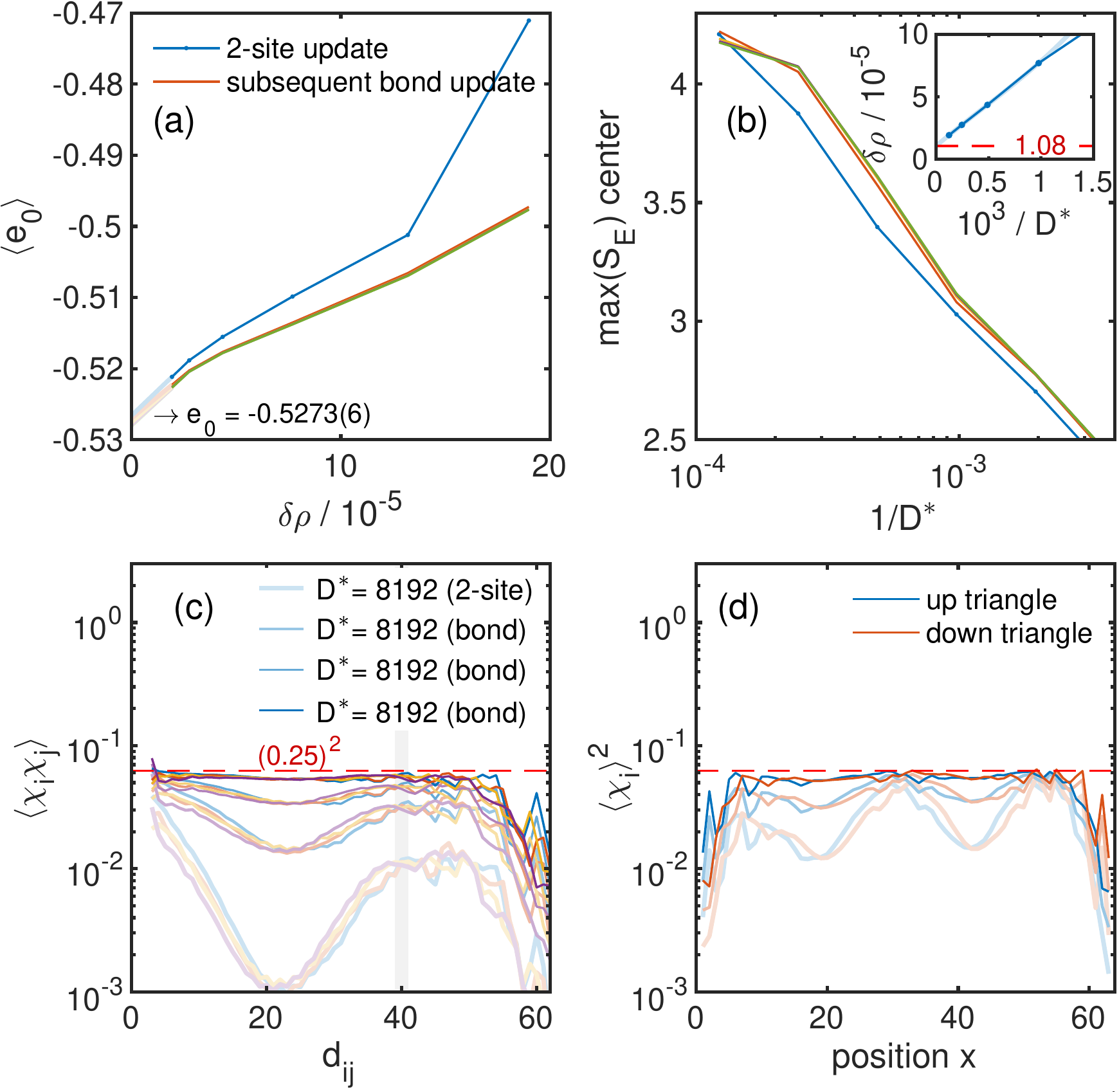}
\caption{{%
   DMRG simulation (complex valued)
   on YC$6\times 64$ at $U=9$. 
(a)-(c) Identical analysis as in \Fig{fig:YC6:real}.
(d) Having complex arithmetic, this permits a non-zero
   expectation value for the chiral order parameter
   $\langle \chi_i\rangle$, showing the square 
   for direct comparison with (c). We note that
   $\langle \chi_i\rangle$ has the same sign
   for all triangles and, consistent with (c),
   converges towards $|\langle \chi \rangle|
   \approx 0.25$ (horizontal red dashed line).
   The data in (d) correspond to the last three
   entries in the legend in (c).
}} 
\label{fig:YC6:cmplx} 
\end{figure} 
\subsection{Degeneracy in the entanglement spectrum}

In addition to the long-range chiral correlation, 
we observe systematic large degeneracy in the
entanglement spectra (ES) defined by $\mathcal{E}_i\equiv-\ln(\lambda_i)$, 
with $\lambda_i$ the Schmidt spectrum of the 
reduced density matrix of half the system.
As shown in \Fig{Fig:Chiral}(c), 
the spectrum levels are grouped versus 
the charge quantum number $\Cz$ 
(relative to half-filling) of the subblock,
and color-coded based on the
spin quantum number $S$. 
The levels are symmetric for $\Cz \to -\Cz$ because the ES are obtained in the system
center. There due to the mirror symmetry, removing (adding) a particle from the
left subblock necessarily adds (removes) it from the right one.

In the infinite-DMRG calculation \cite{Szasz20}
the ground state on the YC4 system is found 
in the semion sector of gapped CSL{, 
where a degeneracy of 2 is observed due to the free spin-\sfrac{1}{2} edge mode}.
Our finite-size DMRG simulations ultimately also 
lead to the same conclusion based on the ES structures
in \Fig{Fig:Chiral}(c-d) in the chiral regime at $U=10$.
However, by comparison to Ref.~\cite{Szasz20},
we find even larger 
ES degeneracies.
In \Fig{Fig:Chiral}(c) the levels show at least a 
8-fold degeneracy. For example, there is systematic 
grouping of two singlets with two triplets
(two blue and red dots, respectively).
A factor $2$ of this degeneracy is
due to our wavefunction being real 
while the system is spontaneously TRS-broken. 
This yields the systematic doubling of {\it any}
spin multiplet in \Fig{Fig:Chiral}(c), 
which can be precisely reduced by conducting 
the same simulation with complex arithmetic 
in \Fig{Fig:Chiral}(d). Correspondingly, an
approximately $\ln(2)$
reduction of $S_E$ is also seen in \Fig{Fig:Obs2}(c).

{In addition to this two-fold degeneracy, our} remaining 
four-fold degeneracy between $S{=}0$ and $1$ can be 
understood as a consequence of the SU(2) DMRG 
simulation on the state with $S{=}\sfrac{1}{2}$ edge 
spinons, similar to the Haldane phase with $S{=}\sfrac{1}{2}$ 
edge modes in the open spin-$1$ chain~\cite{Li2013}.
When computing the ES, we cut the system into two halves
such that additional fictitious edge 
$S{=}\sfrac{1}{2}$ 
degrees of freedom appears at the sub-block boundary.
This leads to a direct product of the
$S{=}\sfrac{1}{2}$ 
edge spinon with the fictitious
$S{=}\sfrac{1}{2}$, 
and thus it gives rise to the sum of a singlet and a triplet, 
i.e., $\frac{1}{2}\otimes\frac{1}{2} \equiv 0 \oplus 1$.
Similarly, with a boundary $S=\sfrac{3}{2}$ excitation,
one arises at $\frac{3}{2} \otimes \frac{1}{2} \equiv 1 \oplus 2$
[red and purple dots in \Fig{Fig:Chiral}(c-d)]. 
The degeneracy of the ES levels 
agrees with the obtained ground state in the semion sector \cite{Wu2020}. 
Importantly, we find
that the low-lying $0\oplus1$ levels satisfy
the $(1, 1, 2, 3, \ldots)$ 
near-degenerate counting, which is consistent with the 
SU(2)$_1$ chiral conformal field theory~\cite{cft}
and thus provides further strong support for
the gapped CSL~\cite{li2008}.

\subsection{{Width $6$ Cylinders (YC6)}}
\label{sec:YC6}

{In this section we proceed to YC6 cylinders.
A major incentive to look at YC6 is
the fact that the $120^{\circ}$ phase for large $U$ fits
naturally into YC6, but {\it not} into YC4 (for this reason
YC4 switches into an RVB-like phase for very large
$U\gtrsim 20$, as already pointed out earlier~\cite{Chen2018}).
As we will see, the chiral intermediate phase also
persists in YC6. We take this as strong support for the
existence of the intermediate CSL {potentially
also in the 2D thermodynamic limit}.

While the YC6 simulations are considerably more
challenging, ultimately, we encounter a rather similar
and thus consistent overall picture as for YC4 \cite{Szasz20}.
Once the cylinders just become long enough, we observe clear
long-range chiral correlations. Let us recall:
the YC$4$ cylinders
established long-range chiral correlations when
(i) the cylinder was sufficiently long
[which turned out to be much longer than the circumference,
$L>18$ for $W=4$ in \Fig{Fig:Chiral}(a)],
and at the same time (ii) the relative energy
accuracy was clearly below 1\%. Assuming
that the bulk gap remains about the same, we have to
aim at an even better relative energy accuracy given
the increased number of sites here for YC$6\times 64$,
before one can start expecting to see long-range chiral
correlations. This makes the YC6 calculations much more
challenging. 

Yet as we show below, we succeed to demonstrate 
the buildup and full establishment of long-range chiral 
correlations also for YC6. {This is in full agreement
with \citet{Szasz20}, which based on iDMRG concluded
that there is also a chiral intermediate phase for YC6.
However, our results are in stark contrast, e.g., to
the more recent variational Monte Carlo simulations
\citet{Tocchio21}. While they confirmed a chiral intermediate
phase for $W=4$, they concluded the chiral phase to be absent
in the low-energy regime of $W=6$ and thus also
in the 2D limit. Similarly, the thermal simulations
of \citet{Wietek21} argued in favor of a gapless
stripy intermediate phase (or a gap too small to be
detected within their DMRG-based approach of minimally
entangled thermal states on YC4 cylinders).
We speculate that such conclusions are related
to the challenges in the DMRG simulations,
as also clearly encountered here based on the
requirement of large system sizes.
Once under control, however, the spin bulk gap in
the chiral intermediate phase is estimated to be large,
i.e., of order $1$ or similarly of order
$J_\mathrm{eff} \simeq \tfrac{4t^2}{U} \approx 0.44$.
For reference, a finite spin-gap in the TLU at intermediate
Coulomb interaction has also been reported
recently in experimental studies on
$\kappa$-(BEDT-TTF)$_2$Cu$_2$(CN)$_3$ \cite{Miksch21}.}}

{The phase boundaries of the intermediate chiral
phase can be expected to be weakly shifted for YC6
as compared to YC4. Indeed, we do not find evidence
for long-range chiral correlation for $U=10$, as 
used for YC4 in \Fig{Fig:Chiral}(a). This is
consistent with the analysis in \cite{Szasz20},
which also showed for $W=6$ that the upper phase
boundary for the chiral intermediate phase moves
towards slightly lower values, having
$\Uh^{\mathrm{YC6}}\lesssim 10$ just below $U=10$.
Hence we focus on $U=9$ for the YC6 system.}
{In \Fig{fig:YC6:real} we present a DMRG simulation
on the YC$6\times 64$ system using real-valued
arithmetic. This is complemented in \Fig{fig:YC6:cmplx}
by an identical simulation, except that it used
complex-valued arithmetic, which thus permits spontaneous
TRS breaking. In these simulations, for the sake of
efficiency, the number \Dstar of kept multiplets
was ramped up quickly by a factor of $2$ in a
two-site update, followed by four sweeps with a plain
bond-update at the same \Dstar.}

{For the real-valued DMRG simulations on YC6
in \Fig{fig:YC6:real}, 
we plot the ground-state energy versus the discarded
weight $\delta\rho$ in \Fig{fig:YC6:real}(a).
The data converges uniformly, except
that it starts to show an onset towards a stronger
decrease of the ground state energy for the very
smallest $\delta\rho \lesssim 2\times10^{-5}$.
This relates to the fact that for the real-valued
DMRG simulations on YC6 we cannot converge the
chiral long-range correlations [\Fig{fig:YC6:real}(c)],
despite keeping up to $\Dstar = 8192$ multiplets
(corresponding to $D=22\,743$ states).
To be precise, while the chiral correlations
in \Fig{fig:YC6:real}(c) appear to decay perfectly
exponentially over long distances, the correlation
length $\xi$ is not converged, in that it keeps
increasing with increasing \Dstar, i.e.,
chiral correlations become stronger.

The maximum entanglement entropy in the system center
vs. $1/ \Dstar$ is tracked in \Fig{fig:YC6:real}(b),
where its inset relates \Dstar to the respective
discarded weight in the DMRG simulation.
Since \Dstar is ramped up very quickly according to the
DMRG sweeping protocol specified above,
it looks as if $\delta\rho$ extrapolates
to a finite value $\delta\rho_0 \simeq 1.22\times10^{-5}$
for $1/\Dstar \to 0$ (horizontal red dashed line).
This is artificial, of course, owing to the sweeping protocol,
and is attributed to the overall
strong truncation still given that the system
barely started to move into the low-energy chiral regime.
Nevertheless, from a practical point of view,
based on that inset in \Fig{fig:YC6:real}(b),
computed quantities may thus be extrapolated
to $\delta\rho \to \delta\rho_0 > 0$ for
quantitative estimates, rather than $\delta\rho \to 0$.}

{The exponentially decaying chiral correlations
in \Fig{fig:YC6:real}(c) were fitted
by $|\chi_0|^2\, e^{-d_{ij}/\xi}$. The resulting
fitting parameters $|\chi_0|$ and inverse
correlation length $1/\xi$ (slope) are summarized in
\Fig{fig:YC6:real}(d) vs. $\delta\rho$.
This suggests that at the same time as
$1/\xi$ extrapolates to zero
at $\delta\rho \gtrsim \delta\rho_0$
[vertical red dashed line, identical with red
dashed line in inset to \Fig{fig:YC6:real}(b)],
the long-range chiral correlation assume
the finite value $|\chi| \approx 0.25$
(horizontal red dashed line).
Therefore despite the fact that the real
DMRG simulation cannot be converged to
explicitly show long-range chiral correlations,
a careful extrapolation of the data
vs. $\delta\rho \to 0$ (or rather
$\delta\rho \to \delta\rho_0$)
does support the conclusion that the
YC$6\times 64$ cylinder based on real-valued
DMRG simulations is chiral for $U=9$.}

{This can be significantly more
substantiated still by repeating
precisely the same DMRG simulation, yet with
complex arithmetic, with the results
presented in \Fig{fig:YC6:cmplx}.
While the convergence of the ground state
energy in \Fig{fig:YC6:cmplx}(a) looks
nearly identical to \Fig{fig:YC6:real}(a), 
the maximal entanglement entropy $S_E$
around the system center in \Fig{fig:YC6:cmplx}(b)
already starts to level off and converge
for the smallest $1/\Dstar$ (largest $\Dstar$ = 8192).
In particular, the entanglement entropy from the
2-site update (blue) remains already the same
for the smallest $1/\Dstar$
when compared to the subsequent bond updates still
(other solid lines).
In \Fig{fig:YC6:cmplx}(c), now one can explicitly
observe how the long-range chiral correlations
build up for over the last DMRG sweeps,
resulting in long-range chiral correlations
of approximately $|\chi|\approx 0.25$ which is in agreement
with the earlier extrapolation in \Fig{fig:YC6:real}(d).
For the complex DMRG, we can also compute
the chiral order parameter $|\chi|$ directly
for a line of triangles along the cylinder
as shown in \Fig{fig:YC6:cmplx}(d). The
data agrees well in magnitude, and thus is
consistent with the chiral
correlations shown in \Fig{fig:YC6:cmplx}(c).}

{The non-extrapolated DMRG ground state energy
of the complex simulation ($e_0 \simeq -0.52270$)
also agrees well with the real-valued calculation
($e_0 \simeq -0.52229$). The former
is just slightly lower as it can explicitly
make use of the TRS breaking. After simple
linear extrapolation of
the ground state energies vs. $\delta\rho\to 0$,
one obtains $e_0 \simeq -0.5273(1)$.
{On a more conservative level, 
extrapolating $\rho \to \rho_0$
yields $e_0 \simeq -0.5245(6)$ which lowers
the ground state energy still by about 0.34\%.}
Now given the presence of a downward kink in the
convergence of the ground state energy at the smallest
energies reached, as already also seen for YC4, the subsequent
extrapolated energy gain may thus again be attributed
to the presence of a chiral bulk gap.
{Here for YC6, the estimate yields
$\Delta_\chi \cong N\cdot (0.0034\,e_0) \sim 0.69$ with
$N= LW$ the total number of sites.
The estimate for $\Delta_\chi$ here for YC6 is about
a factor of $2.7$ larger as previously obtained
in the same manner for YC4. The difference
may be attributed to the slower sweeping protocol
used in the DMRG for YC4 which thus underestimated
the actual gap there. Importantly,
in the present case the gap estimate compares well
to the charge gap $\Delta_c \simeq 0.88$
explicitly evaluated for YC4 in \App{app:gap:ch}
[cf. \Fig{Fig:SPGap}].}}

\begin{figure}[tb!]
\!\!\includegraphics[width=1\linewidth]{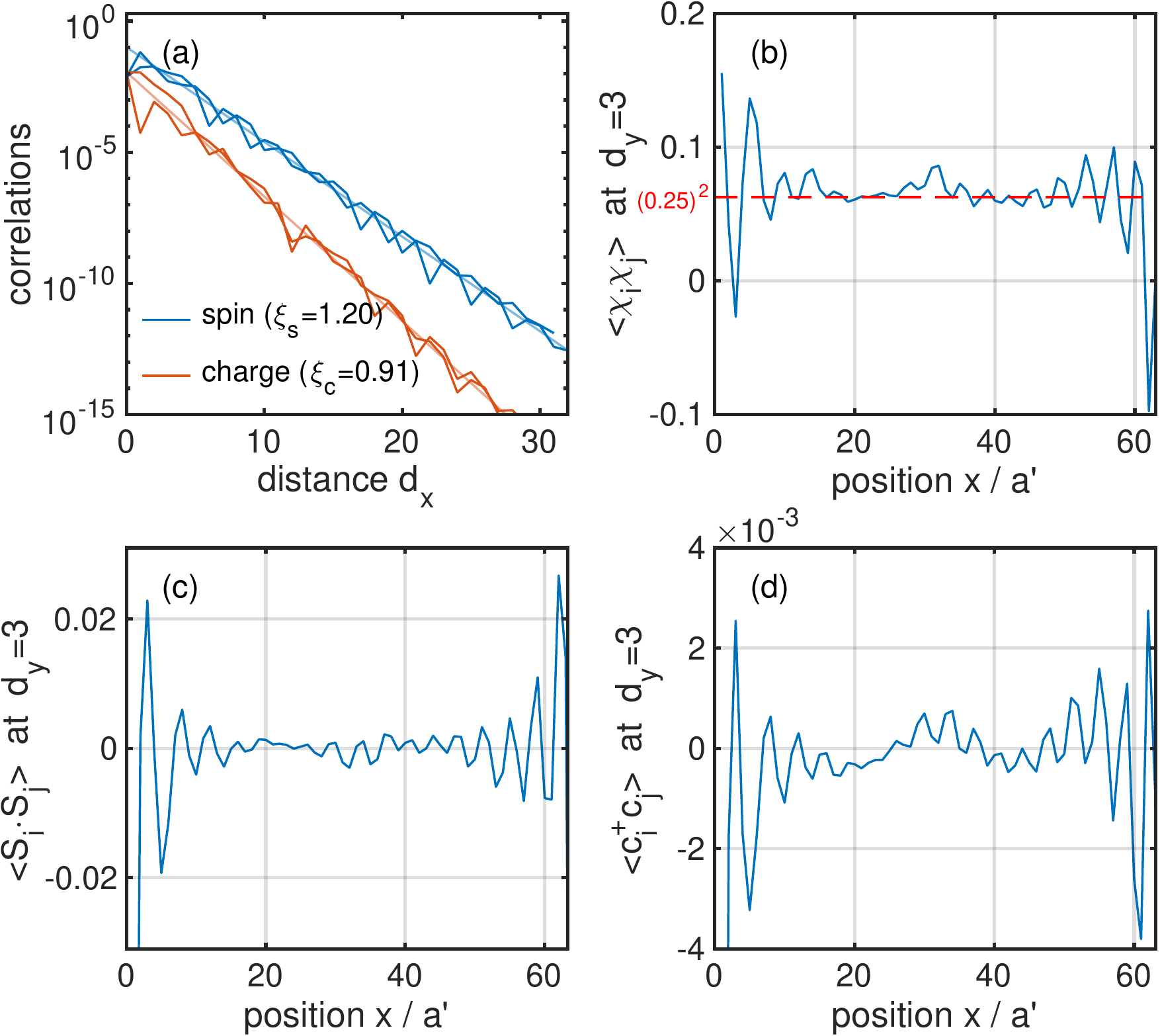}
\caption{{Correlations in the 
YC$6\times 64$ system after the last sweep in the
complex DMRG simulation at $\Dstar{=}8192$
in \Fig{fig:YC6:cmplx} --
(a) Spin $\langle {\bf S}_i {\cdot} {\bf S}_j\rangle$
and charge correlations 
$\langle {\bf c}_{i}^\dagger {\cdot} {\bf c}_{j}\rangle$ (where
the dot product sums over spin)
relative to the system center. The exponential
fit $\propto e^{-x/\xi}$ (straight line in matched light color)
yields the corresponding correlation length
as shown with the legend.
The correlations are computed along a tilted straight
path of length $L-1$ in units of lattice spacing
from left to right YC boundary [cf. \Fig{Fig:Phase}(a)].
(b) Chiral correlations, (c) Spin correlations,
(d) Charge correlations, all at constant vertical distance
$d_y{=}3$ ($j{=}2,5$) vs. position $x$ along the cylinder.
Here $x$ is the distance from the boundary of the cylinder
converted to column index $x/a'\in [1, L]$
[cf. \Fig{Fig:Phase}(a)].
}}
\label{fig:YC6:correl} 
\end{figure} 

\subsection{{On the need for long cylinders}}

{All of the systems above required an a priori
surprisingly large degree of asymmetry in the aspect
ratio of the cylinder geometry in order to realize
the intermediate phase with long-range chiral correlations.
Specifically, this required the cylinders to be much longer
(i.e., in the direction of the open boundary) as compared to
their circumference (periodic boundary).
With the notion in mind that for a topological system,
the bulk gap needs to close towards the boundary,
the finite open boundary of the cylinders studied,
in principle, can affect the cylinder considerably
into the bulk itself. Naively one may have speculated
that the effects of the open edges diminish quickly
like with the bulk correlation lengths for spin
or charge which, based on the data in \Fig{Fig:Obs2}
for YC4, are ultrashort on the order of a single
lattice spacing. But this ignores
the fact, that the bulk gap actually needs to physically
close towards the open boundary. The length scale
over which this occurs is, a priori, far from clear.}

{In order to gain further insight into the effect of
the open boundary, we look more closely into correlations
also with reference to the boundary for YC$6\times64$.
The results are summarized in \Fig{fig:YC6:correl}
where we take the ground state from the complex-valued DMRG
simulation as in \Fig{fig:YC6:cmplx} after the last sweep.
While $\Dstar=8192$ already
ensures visible convergence and established long-range
chiral correlations, the variations in the data
in \Fig{fig:YC6:correl}(b-d)
around the system center $x \sim L/2=32$
need to be taken with a grain of salt, bearing in mind
the residual variations in the data of the last sweep
in \Fig{fig:YC6:cmplx}(d) itself.}

{To start with, the bulk spin and charge correlations
for YC6 remain extremely short ranged, as seen
\Fig{fig:YC6:correl}(a). The correlation lengths
are of about one lattice spacing, consistent with
the YC$4$ data in \Fig{Fig:Obs2}.
Given the slightly smaller $U=9$ here,
the correlation length for charge transfer
is slightly increased, ($\xi_c=0.81 \to 0.91$),
yet the spin-spin correlation length is actually
reduced ($\xi_s=1.66 \to 1.20$), bringing the
two correlation lengths closer to each other.}

{Now in order to analyze effects vs. distance
from the boundary, we compute chiral, spin, and charge correlations
at fixed vertical distance $d_y=3$ of site $i$ relative
to site $j$ (i.e. taken halfway around the cylinder)
vs. position $x / a' =1,\ldots,L$
along the cylinder, with the results
shown in \Fig{fig:YC6:correl}(b-d), respectively.
While superficially, the data behaves similarly,
the magnitudes vary considerably, with the scale of values
decreasing from chiral to spin to charge correlations,
in agreement with the expected respective relevance
in the low-energy regime.
Furthermore, one notices that all data shows pronounced
oscillatory behavior close the boundary with a period
of around four columns in YC6, i.e.,
$\lambda \simeq 4 a'$. More importantly,
the enveloping amplitude decays rather slowly
into the bulk, taking about $\Delta x \sim 5 \lambda$,
i.e., $20$ columns from the open boundary of the YC6
cylinder to diminish.
Together with the right boundary, this suggests that
in order for the effects of the open boundary to 
have significantly decayed to actually see bulk
behavior in the system center,
one needs a rather long cylinder with
$L \gtrsim 10\lambda \sim 40$ columns.
For shorter systems, the two
boundaries can thus be expected to interfere with
each other which can be detrimental to the
development of long-range chiral correlations,
as observed, for example, for the shorter YC$4$ systems
in \Fig{Fig:Chiral}.}

\section{Summary and Conclusions}
\label{sec:summary}

{{We show clear numerical evidence for an
intermediate chiral spin-liquid phase on long yet
finite-size YC4 and YC6 cylinders of the half-filled triangular
lattice Hubbard model for sizable $U \in [U_{c1},U_{c2}]$
based on exact large-scale DMRG simulations.
This phase is surrounded by}
a metallic phase for $U<U_{c1}$ (or possibly
a Luther-Emery liquid \cite{Gannot20,Szasz21}),
and a $120^{\circ}$ magnetic insulating phase
for $U>U_{c2}$.
For YC4
we find $U_{c1} \simeq 9$ and $U_{c2} \simeq 10.75$,
whereas for YC6, $U_{c2} \lesssim 10$.
The intermediate spin-liquid phase has been debated
intensely in recent literature, with 
contradicting conclusions on whether it represents
a chiral spin liquid or not.}
Our results demonstrate that finite-size
effects can drastically alter the conclusions.
Here, the system length in DMRG simulation constitutes
a key factor to identify the 
spontaneous TRS breaking in the CSL state.

In the effective spin model derived from the Hubbard model
\cite{Motrunich2005,Sheng2009,Yang2010,Cookmeyer21},
the ring-exchange couplings have order $t^4 / U^3$.
Previous study estimated the metal-insulator 
transition to occur at $U / t \simeq 5$~\cite{Morita2002}, 
leading to important ring-exchange couplings that can 
drive a spinon Fermi surface state~\cite{Motrunich2005}.
However, DMRG studies find the Mott transition at 
$U / t \simeq 9$, which indicates the much weaker 
ring-exchange couplings and may explain
why the spinon Fermi surface state is not found.
{Properly accounting for 
spin couplings 
in an effective spin model allows one
to understand the emergence of the gapped CSL
of Kalmeyer-Laughlin type \cite{Cookmeyer21}.}

{A handwaving argument
on a possible origin of the CSL phase at the intermediate 
$U$ may be taken from the phase diagram a finite 
temperature \cite{Wietek21}}. In the large $U$ limit, TLU 
reduces to an effective Heisenberg spin model. Besides the 
long-wavelength soft modes at the $K$ point corresponding 
to the $120^\circ$ order, there exists additional rotonlike modes 
near the $M$ point at higher energy~\cite{Chen2019}.  
The softening of the $M$-point rotonlike excitations,
either by quantum 
fluctuations~\cite{Ferrari2019} or thermal fluctuations~\cite{Chen2018,Chen2019}, 
seems to be accompanied with an emergent liquid-like phase with anomalously 
enhanced chiral fluctuations. Here the charge fluctuations also lead to a spin liquid phase
with long-range chiral order and enhanced spin fluctuations at the $M$ point,
naturally implying the softening of the rotonlike excitations 
and spinon deconfinement in the transition with decreasing $U$.
Therefore it would be interesting and important to explore the spin dynamics
of this Hubbard model in future study.

\begin{acknowledgments}

We acknowledge the stimulating discussions with Olexei I. Motrunich and Hong-Hao Tu.
W.L., Z.C., and S.-S.G. were supported by the National Natural Science Foundation of 
China grants No. 11974036, No. 11834014, No. 11874078, 583 No. 12074024, No. 11774018, No. 12222412, 
the Fundamental Research Funds for the Central Universities, 
and CAS Project for Young Scientists in Basic Research (Grant No. YSBR-057).
D.N.S. was supported by National Science Foundation Grant PREM 
DMR-1828019. B.B.C. was also supported by the German Research foundation, DFG WE4819/3-1. 
B.B.C. and W.L. thank the High-performance Computing Center at ITP-CAS for the technical support 
and generous allocation of CPU time. 
A.W. was supported by the U.S. Department of Energy,
Office of Science, Basic Energy Sciences,
Materials Sciences and Engineering Division.

\end{acknowledgments}

\begin{appendices}

  \setcounter{figure}{0}
  \renewcommand{\thefigure}{A\arabic{figure}}
  
\section{Symmetric Construction of Chiral Operator}
\label{app:chirop}

\begin{figure}[thb]
\includegraphics[width=0.95\linewidth]{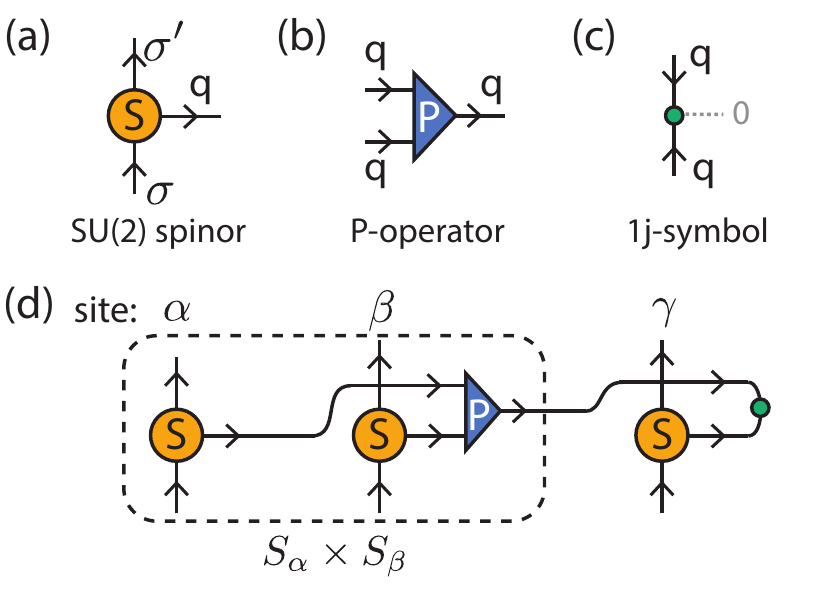}
\caption{
Symmetric tensor representation of 
(a) SU(2) spinor $S^q$ [\Eq{Eq:Spinor}], (b) $P$-operator, (c) the 1j-symbol, 
and (d) the scalar chirality operator order $\chi$ [c.f. \Eq{Eq:Chi}]. 
}
\label{Fig:ChiOp}
\end{figure}

With spin rotation SU(2) symmetry implemented, 
the spin operators need to
be reorganized into an irreducible operator (irop),
i.e., the spinor also schematically
depicted in \Fig{Fig:ChiOp}(a) \cite{Weichselbaum2012},
\begin{equation}\label{Eq:Spinor}
\hat{S}^{q=1,m} \equiv \begin{pmatrix} 
   \frac{-1}{\sqrt{2}} \hat{S}^+ \\[1ex]
   \hat{S}^z \\
   \frac{1}{\sqrt{2}} \hat{S}^-
   \end{pmatrix}, 
\end{equation}
with the ladder operators $S^+=(S^x+iS^y),~S^-=(S^x-iS^y)$.
Here the components 
$\hat{S}^{1,+1}$, 
$\hat{S}^{1,0}$, 
and $\hat{S}^{1,-1}$, 
transform like an
irreducible representation (irep) 
$|S; S_z\rangle$ with $S_z=+1,0,-1$, respectively.
For this, the relative sign on the first component
is important.
The spin operator always corresponds
to the adjoint representation, i.e., $S=1$ for SU(2).
The normalization of the irop in \Eq{Eq:Spinor}
is chosen such that $\hat{S}_i^\dagger \cdot
\hat{S}_j^{\,}$ corresponds to the standard
Heisenberg interaction $\hat{S}_i \cdot \hat{S}_j$
where, nevertheless, when having \Eq{Eq:Spinor},
the dagger on $\hat{S}_i$ becomes important.

Next we can build the scalar chirality,
which we expand as 
\begin{subequations}
\begin{eqnarray}
  \hat{\chi}_{ijk} &=&
(\hat{\sigma}_i \times \hat{\sigma}_j) \cdot \hat{\sigma}_k \equiv
8\, (\hat{S}_i \times \hat{S}_j) \cdot \hat{S}_k
\label{Eq:Chi:1} \\[1ex]
   &=& 4 \iu\,\Bigl(
      \hat{S}_i^+ \hat{S}_j^- \hat{S}_k^z + 
      \hat{S}_i^z \hat{S}_j^+ \hat{S}_k^- + 
      \hat{S}_i^- \hat{S}_j^z \hat{S}_k^+ 
      {\ - \  {\rm H.c.}}
    \Bigr)
. \label{Eq:Chi}
\end{eqnarray}
\end{subequations}
Now while this may look somewhat tedious, in practice,
there is a simple transparent procedural way for dealing
with it {from a tensor network perspective with SU(2) spin
symmetry enabled. As required for an observable,
the chiral operator is a scalar operator
with nonzero eigenvalues $\pm \sqrt{12}$}. 
It combines three spin operators $\hat{S}$ into a scalar 
operator $\hat{\chi}$. As depicted in \Fig{Fig:ChiOp}(d),
one needs to ``tie together'' the three $S=1$
irop indices (horizontal leg for each $S$).
This can be simply achieved by fusing two $S=1$
multiplets (ingoing) into $S=1$ (outgoing).
From a symmetry perspective, this is also the
only possible combination here.
The result corresponds to a tensor $\hat{P}$ that
is proportional to the Clebsch-Gordan
coefficient tensor (CGT) $C_{1,1}^1
\equiv (1,1|1) \propto P$
[\Fig{Fig:ChiOp}(b)].
With this, the chiral term can be compactly written
as the nested contractions {(denoted by $\ast$)}
\begin{eqnarray}
   \hat{\chi}_{ijk} = \Bigl(
     \hat{S}_i \ast \bigl( \hat{S}_j \ast \hat{P} \bigr)
   \Bigr)
   \ast \hat{S}_k^\dagger
\text{ ,}
\end{eqnarray}
with the precise order of pairwise contractions
by the brackets irrelevant.
Optionally, this expression can be further
symmetrized. The direction of an arrow (leg)
can be reversed based on a so-called `1j' symbol
\cite{Weichselbaum2020}. In the present case
this corresponds to the CGT $(1,1|0) = \tfrac{1}{\sqrt{3}} U$
that fuses two $S=1$ multiplets into a singlet.
The index with the outgoing $S=0$ is a singleton dimension
and hence can be skipped. After proper normalization,
this reduces to a unitary operator $U$ [\Fig{Fig:ChiOp}(c)].
Thus inserting $U^\dagger\ast U = 1$,
having $U=U^\dagger$ here, and
contracting one $U$ onto the outgoing index of $P$,
denoted as $P_A\equiv P\ast U$
[green dot contracted onto blue triangle in \Fig{Fig:ChiOp}(d)], 
and the other $U$ onto $S_k^\dagger$ (which
effectively removes the dagger),
the chiral term becomes [\Fig{Fig:ChiOp}(d)]
\begin{eqnarray}
   \hat{\chi}_{ijk} = \Bigl(
     \hat{S}_i \ast \bigl( \hat{S}_j \ast\hat{P}_A \bigr)
   \Bigr)
   \ast \hat{S}_k
\text{ ,}
\end{eqnarray}
now with all three spin operators $\hat{S}$ on an
equal symmetric footing.
Here $P_A$ is a completely antisymmetric tensor 
for its three indices, all of which are incoming now.
As such, it corresponds to a Wigner 3j symbol,
that in the present case precisely corresponds
to the Levi-Civita tensor as it appears in
the original definition of the chiral operator
in \Eq{Eq:Chi:1}, up to an overall purely imaginary
normalization factor.

\section{Convergence of the DMRG calculation}
\label{app:convergence}

\begin{figure}[thb]
\includegraphics[width=0.99\linewidth]{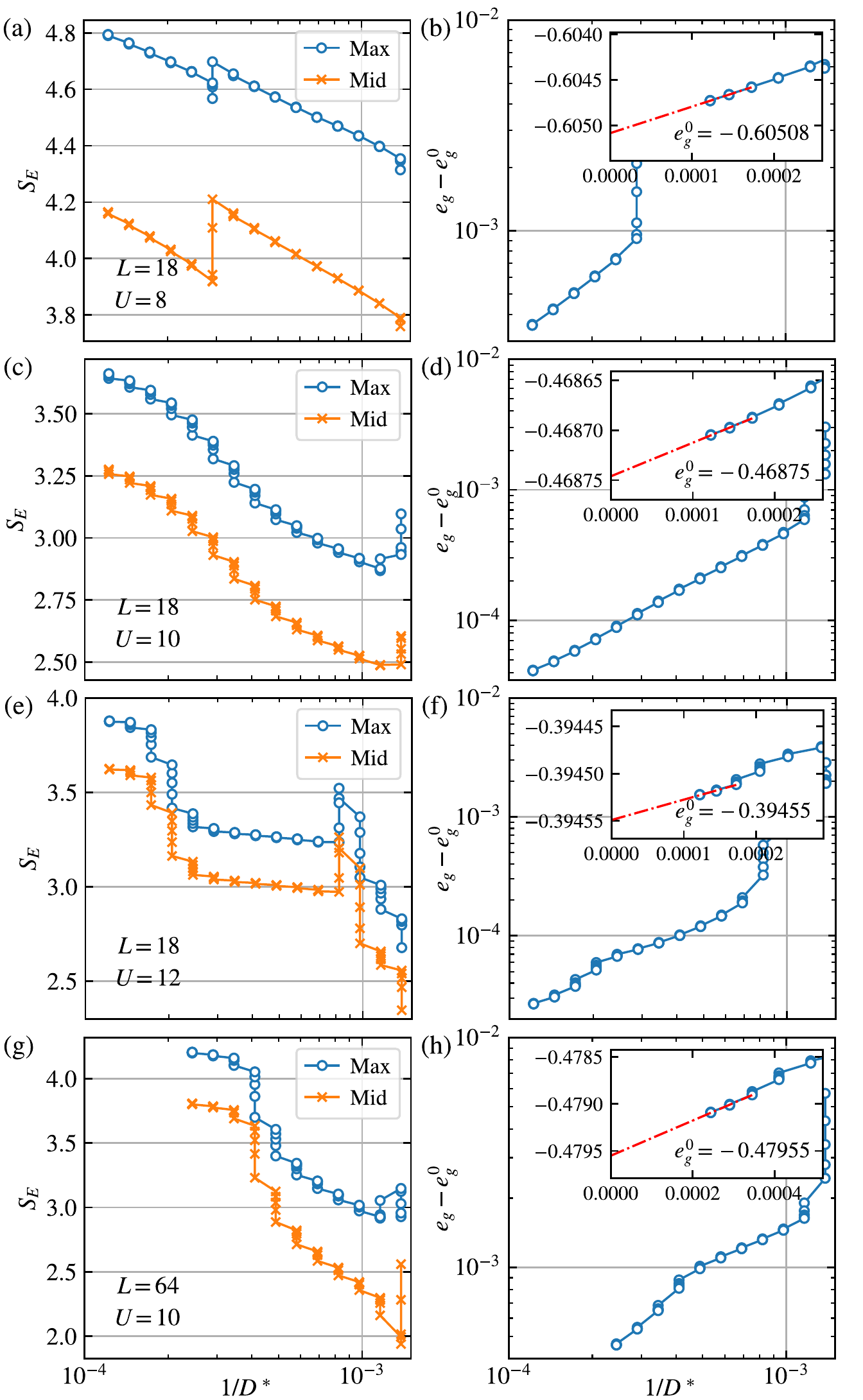}
\caption{
Entanglement entropy $S_E$ (left panels) and ground-state energy (right panels) vs. 
$1/D^\ast$ for YC4 for (a, b) $U = 8$; (c, d) $U=10$; (e, f) $U=12$ at $L=18$; and for (g, h) $U=10$ 
at $L=64$. Entanglement entropy $S_E$ labeled 'Max' 
are the maximal entanglement amongst all bonds throughout the system; 
while the 'Mid' ones are values measured at the center of the systems. 
The linear extrapolations with $1/D^\ast$ are shown in the inset of all right panels (b,d,f,h).
Although the calculations are very challenging in the 
metallic phase with $U=8$, the data have reached very good convergence for 
$U=10$ and 12. 
}
\label{Fig:Conv}
\end{figure}

In this section, we show the computed entanglement entropy $S_E$ and 
ground-state energy per site $e_g$ vs. bond dimension $1/\Dstar$. 
As can be seen below, the results shown in the main text for the 
intermediate chiral spin liquid (CSL) and large-$U$ regimes are well-converged vs. \Dstar.
In practice, to ensure convergence of the data, 
we ramp up the bond dimension \Dstar{}
for sweep $n$
in the {uniform exponential manner, as described by} 
$\Dstar_n = \Dstar_0 \cdot a^n$. 
Here $a$ is a parameter that controls the speed of 
increase of \Dstar from one sweep to the next,
until the final bond dimension $\Dstar_{n_\mathrm{max}}$ 
is reached.
We start with an initial bond dimension
$\Dstar_0$ during a random initialization in
the global symmetry sector $(S,Q)=(0,0)$
where $S$ and $Q$ denote
the spin and charge quantum numbers, respectively.
In this work, we use $\Dstar_0=512$, 
$a=2^{1/5}$, 
$\Dstar_{n_\mathrm{max}}=8192$ 
and thus $n_\mathrm{max}=20$ sweeps.
This leads to well converged results in most cases. 
In addition, we may slow down the ramping
of \Dstar by actually performing
up to $5$ sweeps for a given `stage' $n$,
before moving on to the next
stage $n+1$ with increased bond dimension
$\Dstar_{n+1}$, e.g., seen as vertical stacking of data points
in \Fig{Fig:Conv}.

In the left panels of \Fig{Fig:Conv},
we show particular specifics
of the block-entanglement entropy
simulations as we ramp
up the number of multiplets $D^\ast$
in our DMRG simulations. We
show both the maximum values of entanglement entropy $S_E$ 
(labeled by `Max') and the ones cutting at the center bond of the system (labeled by `Mid').
In \Fig{Fig:Conv}(a), for $U=8$, we can see the fast growth of entanglement entropy $S_E$, 
expected in metallic state that is extremely challenging for the DMRG calculation, 
even though overall convergence, e.g. of the ground state energy, already appears systematic.
For the CSL phase at $U=10$ [panel (c,g)] and the magnetically ordered phase at $U=12$ [panel (e)], 
$S_E$ results are well converged vs. $1/\Dstar$.

In the right panels of \Fig{Fig:Conv}
we check the convergence of the ground-state energy $e_g$.
We linearly extrapolate it towards $1/\Dstar \to 0$
based on the last three data points (red line in insets).
We use the extrapolated value
$e^0_g \equiv \lim_{1/\Dstar\to0} e_g(\Dstar)$,
to estimate the `error' of finite-bond-dimension energy 
as the difference $\epsilon_e \equiv e_g(\Dstar) - e_g^0$,
as shown in the main right panels in \Fig{Fig:Conv}. 
From these we see that, within their respective system size, 
at $L=18$ the energies are converged to
$\epsilon_e {\sim} {10^{-4}}$ for $U=8$ 
and to $\epsilon_e {\sim} {10^{-5}}$ 
for $U=10,12$; and at $L=64$, the energies are converged to 
$\epsilon_e {\sim} {10^{-4}}$ for $U=10$.

Overall, convergence is not always smooth
with increasing bond dimension \Dstar. For example,
there may be excitations in the system due to the
arbitrary initialization of the wave function
for small \Dstar which, nevertheless, get ironed out early on.
Certain low-energy excitation as well as edge modes
can be dealt with over longer distances
only once a sufficient accuracy, i.e., sufficiently
large \Dstar, has been reached
which then may lead to a rather sharp drop or increase
in the maximal block entanglement across the entire system, 
as well as a rapid drop of energy, as observed 
for intermediate and also larger \Dstar.

On a physical level, a certain
choice of \Dstar permits a certain energy resolution,
as evident from the analysis of the right panels
of \Fig{Fig:Conv}. This gives
insights into the energy scales of the system
under consideration. Consider, for example,
the bottom panels of \Fig{Fig:Conv}
for the chiral intermediated state at $U=10$
for $L=64$. There the chirality of the ground state
only emerges for $\Dstar \gtrsim 2\,500$
($1/\Dstar \lesssim 4
{\times}
10^{-4}$)
which leads to a sharp rise in the entanglement
entropy [\Fig{Fig:Conv}(g)]. This by itself already
suggests a significant change in the underlying
DMRG wave function, e.g., as also seen
in the convergence of the chiral
long range correlation in Fig. 4(a) of main text.
Indeed, the systematic degeneracies in the
entanglement spectra also only emerge once
the sufficiently large $\Dstar \gtrsim 2\,500$
is reached in the present case.
Simultaneous with strong rise in the entanglement
entropy in \Fig{Fig:Conv}(g),
one also observes a kink in the convergence
of the ground state energy [\Fig{Fig:Conv}(h)].

\begin{figure}[thb]
\includegraphics[width=1\linewidth]{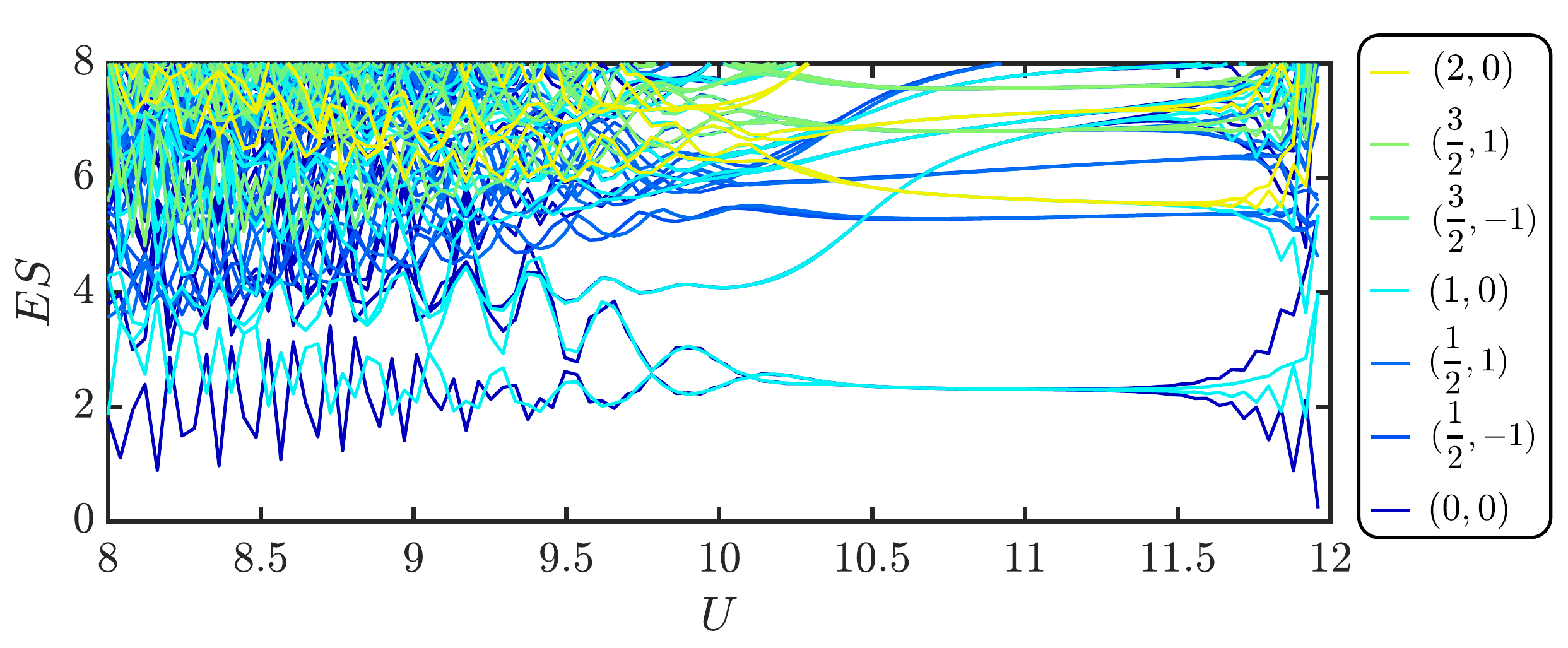}
\caption{Entanglement spectrum (ES) along the
system and in between full columns of a DMRG scan
along a a YC4$\times$100 cylinder. For this
we tune $U\in[8,12]$ 
linearly along the cylinder. We keep the bond dimensions
up to $\Dstar = 5793$ ($D \lesssim 17053$).
The lines are color coded according their symmetry
sector $(S, \Cz)$ as indicated.
}
\label{Fig:ES}
\end{figure}

This suggests that for the chiral state to be
seen in the DMRG simulations, one needs an energy
resolution as found at the kink, that is
$\Delta e_\chi \cong e_g - e_g^0 \simeq 10^{-3}$.
If one were to interpret this to reflect an
actual energy gap below which the DMRG convergence
is accelerated, the corresponding estimate
would be $\Delta_\chi \cong
LW\Delta e_\chi \simeq 0.25$ and thus sizable.
Increasing the ground state energy
of the $L=64$ system by adding $\Delta e_\chi$,
it still has a significantly
lower energy than the $L=18$ system, given that
$\Delta e_g^0 = e_g^0(L=18) - e_g^0(L=64) \sim 0.011$.
Conversely then, a non-chiral state may be seen 
if the energy of a given (eigen) state is 
higher by $\Delta_\chi$ above the
ground state for given width $W=4$ system,
either due to insufficient \Dstar or
due to finite-length effects. The latter
is demonstrated by comparison to the same
$U=10$ Hamiltonian, yet for the significantly shorter
$L=18$ in \Fig{Fig:Conv}(c) and (d). From the above estimates,
the finite size correction for this smaller system
size is significantly larger than $\Delta_\chi$,
and so we do not yet see a chiral signatures in
its converged ground state
[see also \Fig{Fig:SGap}(a)]. Therefore also
the entanglement profile in \Fig{Fig:Conv}(c)
evolves much more smoothly as compared to
\Fig{Fig:Conv}(g), and the convergence of the
ground state energy in \Fig{Fig:Conv}(d)
also shows no kink.

\begin{figure}[thb]
\includegraphics[width=1\linewidth]{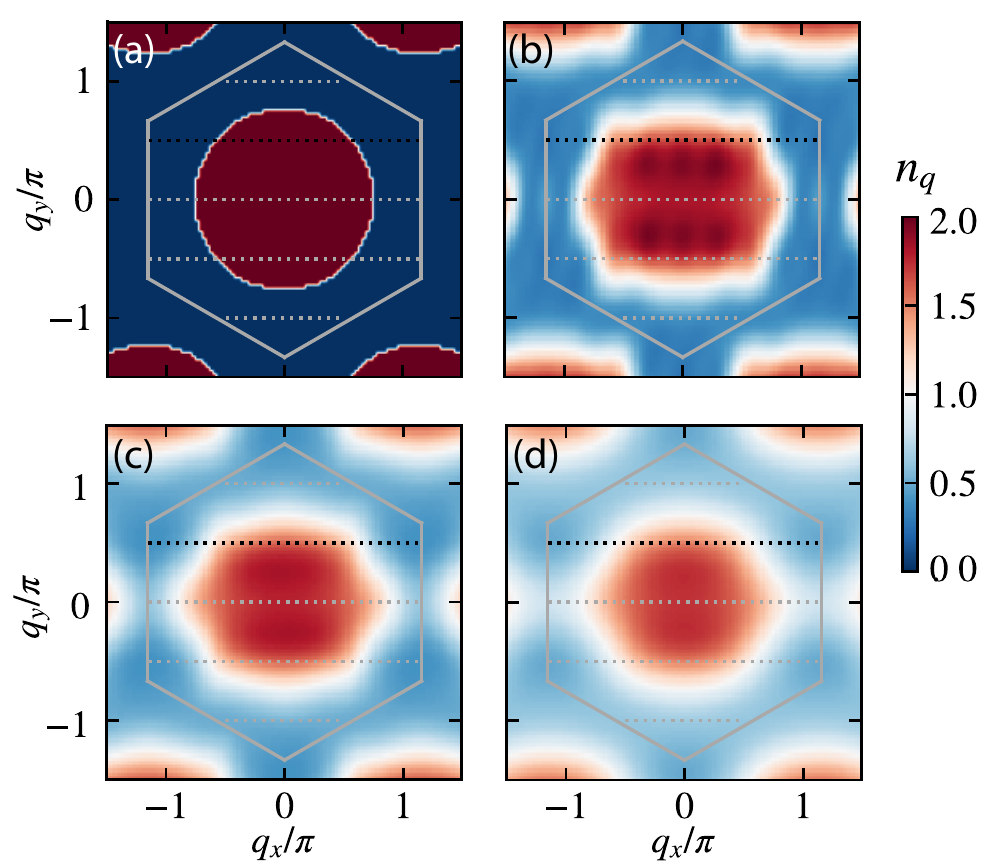}
\caption{Electron density in the momentum space
$n_q$ on the YC4$\times$18 systems with 
(a) $U$=0 (b) $U$=8 (c) $U$=9 (d) $U$=10. 
Grey hexagons represent the boundary of first Brillouin zone, 
and the horizontal dotted lines represent the allowed momenta $q_y$ for the YC4 cylinder.
}
\label{Fig:nk}
\end{figure}

\section{DMRG scan of the cylinder with smoothly changing $U$} 
\label{app:DMRG:scan} 

To study the phase diagram of the model,
we also performed a linear DMRG scan \cite{Zhu2015} 
for a YC4$\times$100 system with the varying $U$ for different
columns $i$ as
$U_i = U_1 + \frac{i-1}{L-1} (U_L-U_1)$. 
Here, we set $U_1=8$ for the first column and $U_L=12$ for the last one.
The results are shown
in \Fig{Fig:ES}, where we compute and collect
the entanglement spectra 
for each cut between the columns $i$ and $i+1$. Since each column $i$ corresponds 
to a unique $U_i$, the ES `flows' as $U$ changes
along the cylinder. 

In the present YC4 system, three possible phases can be 
discerned in the entanglement spectrum, separated by the
two critical points at 
$\Ul \simeq 9.5$ and $\Uh \simeq 10.5$. 
The ES is non-degenerate in the small-$U$ ($U<9.5$) phase,
while it approaches $4$- and $8$-fold degeneracy 
in the intermediate-$U$ ($9.5<U<10.5$) and
large-$U$ ($U>10.5$) regime
(2- and 4-multiplet degeneracy, respectively). 
When approaching the open right
boundary of the cylinder at $U=12$,
the degeneracy starts to split again.
This is consistent with the splitting in the ES 
between the $S=0$ and $S=1$ multiplets
already discussed with Fig. 4(c-d) in main text.

\section{Destruction of the Fermi surface}  
\label{app:Fermi:surface}

In \Fig{Fig:nk}, we show the electron density in momentum space 
$n_q$ on the YC4$\times$18 system. 
The distribution $n_q$ of the free fermion system at $U=0$ in the thermodynamic limit
is also included in \Fig{Fig:nk}(a) as a reference, which shows a perfect Fermi surface.
In \Fig{Fig:nk}(b)-(d), the calculated $n_q$ exhibits a recognizable 
Fermi surface at $U=8$. With further growing $U$, the Fermi surface gets significantly blurred at 
$U=9$ and $10$, characterizing a metal-insulator transition with destructed Fermi surface.

\section{Charge gap}
\label{app:gap:ch}

\begin{figure}[t!]
\includegraphics[width=0.95\linewidth]{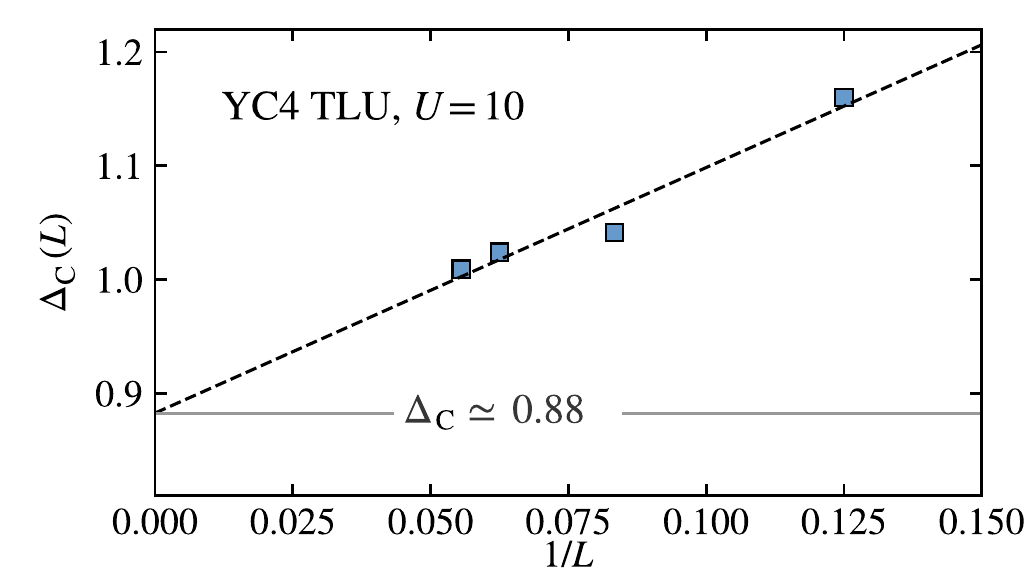}
\caption{Charge gap $\Delta_\mathrm{C}$
calculated on the YC4 cylinders with different lengths $L=8, 12, 16, 18$. 
The linearly extrapolated value with $L\to\infty$ is $\Delta_\mathrm{C}\simeq0.88$. 
}
\label{Fig:SPGap}
\end{figure}

In this section, we directly calculate the
charge (single-particle excitation) gap,
\begin{eqnarray*}
  \Delta_\mathrm{C}
  = \tfrac{1}{2}[E(\tfrac{1}{2}, {+1}) + 
E(\tfrac{1}{2}, {-1}) - 2E(0, 0) ],
\end{eqnarray*}
where $E(S, {\Cz})$ 
denotes the lowest eigenenergy with total spin $S$,
and charge {$\Cz$ taken as the number of particles
relative to the half-filling. 
For the ground state we have $S=\Cz=0$.}
As shown in \Fig{Fig:SPGap} for $U = 10$, 
$\Delta_\mathrm{C}$ decreases with system length $L$, 
and the linear extrapolation over $1 / L$ results in
a large nonzero gap $\Delta_\mathrm{C}\simeq 0.88$.
This confirms that the spin liquid resides in the 
Mott insulating phase.

\section{Identification of edge spinons from spin excitation} 
\label{app:edge:x}

\begin{figure}[t!]
\includegraphics[width=.9\linewidth]{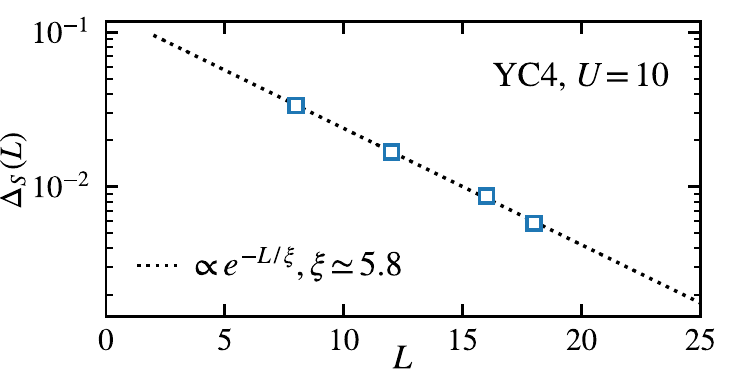} 
\caption{Energy difference $\Delta_S$ between the total spin-0 and total spin-1 sectors.
We calculate the YC4 cylinders with system length $L=8, 12, 16, 18$. 
This shows an exponential decay with $\xi\simeq 5.8$ 
with increasing $L$ as indicated. 
}
\label{Fig:SGap}
\end{figure}

\begin{figure}[t!]
\vspace{.1in}
\includegraphics[width=.95\linewidth]{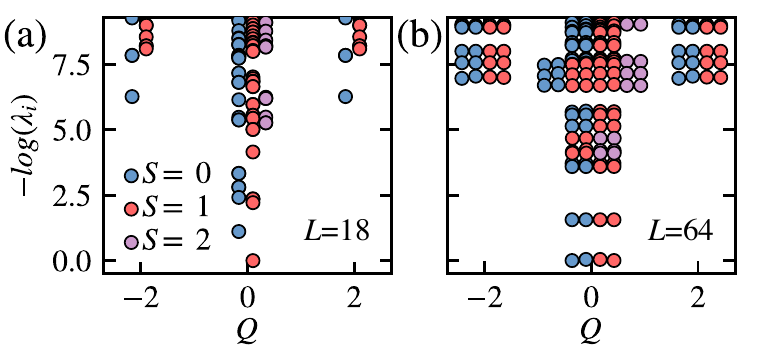}\vspace{-.1in}
\caption{
Entanglement spectrum calculated at the center of YC4 systems with 
(a) $L=18$, and (b) $L=64$ [cf. Fig. 4(c) in main text],
grouped by charge sectors (showing even \Cz only), 
with spin labels color coded as specified in the legend.
\vspace{-.1in}
}
\label{Fig:ESGap}
\end{figure}

In the CSL phase, we have found the obtained ground state (on the YC4 cylinder) in 
the semion topological sector, with a spin-$1/2$ spinon on each open boundary. 
Therefore we expect a 4-fold ground state
degeneracy ($S=0 \oplus 1$) for sufficiently long cylinders.
This may be seen analogous to
the Haldane phase of the open $S=1$ spin chain,
where also spin-$1/2$ edge modes are weakly coupled.
Thus, the energy difference between the total spin-0 and spin-1 sectors
is expected to decay exponentially 
with growing system size~\cite{Kennedy1990}.

Here, we use the similar strategy to identify the edge spinons in the semion sector of the CSL state. 
We show the energy difference $\Delta_{S}$ between the total spin-0 and spin-1 sectors with growing
system length in \Fig{Fig:SGap}. Clearly, $\Delta_{S}$ is very small and indeed decays exponentially with $L$. 
Since the spin triplet excitation is gapped in the bulk
(c.f. \Fig{Fig:Obs2} in main text), this vanishing energy 
difference must be ascribed to the edge spinon modes in the semion sector of the CSL. \\

\section{Finite-size effect of TRS breaking detection}
\label{app:finite-size}

To further emphasize
the absence long-range chiral correlation on short systems, 
we also contrast the ES for $L=18$ to $L=64$ in the main paper.
As seen in \Fig{Fig:ESGap}, 
$L=18$ is still clearly qualitatively different.
An obvious difference from the ES for $L=64$
[\Fig{Fig:ESGap}(b)] is
the complete absence of degeneracies
which, in particular, demonstrates 
that the two low-lying real wave functions that
respect TRS are still split by 
a relatively large gap due to finite-size effects.
Therefore simulations on such short systems may 
lead to the premature conclusion of no chiral order. 
However, with growing system length $L$, 
this gap decreases, and eventually becomes negligible.
This then allows DMRG calculation to obtain the minimal 
entangled state with spontaneous TRS breaking~\cite{jiang2012}.

\section{Chiral correlation on XC4 system}
\label{app:XCdata}
\begin{figure}[htp]
\vspace{.1in}
\includegraphics[width=.95\linewidth]{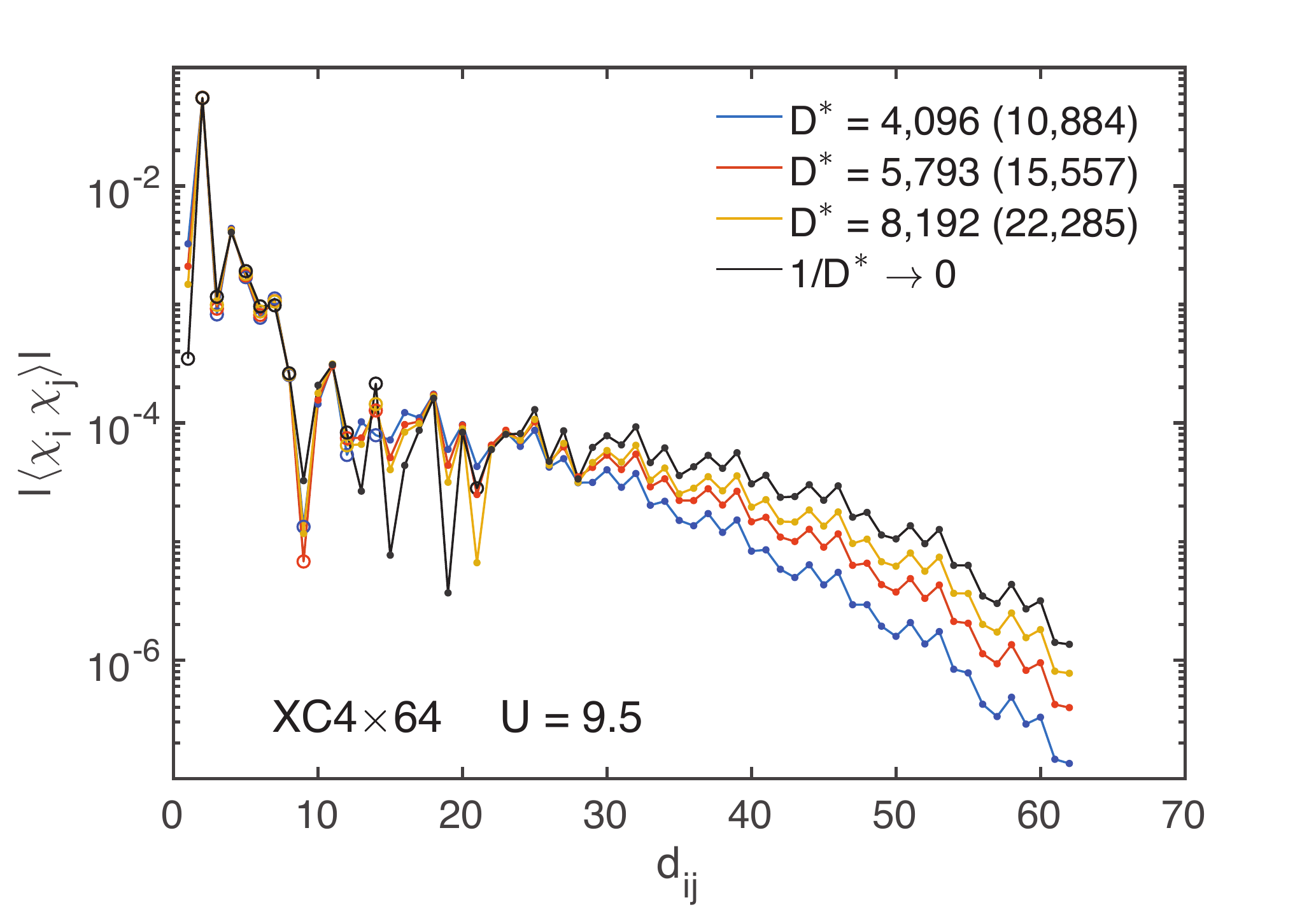}\vspace{-.1in}
\caption{
{Chiral correlations calculated in a $U=9.5$ XC4$\times$64 system with bond dimensions 
$D^\ast=4096, 5793, 8192$, show no long-range correlations. The black line depicts the 
results of linear extrapolation $1/D^\ast\to0$ from the above bond dimensions. 
Here, the filled symbol indicated the positive sign of
chiral correlation, and otherwise the sign is negative.}
\vspace{-.1in}
}
\label{Fig:XCChiral}
\end{figure}

{
In this section, we show the chiral correlations $\langle \chi_i\chi_j \rangle$ between two triangles 
labeled $i$ and $j$ with distance $d=|i-j|$, in XC4$\times$64 systems for the case of $U=9.5$. 
As shown in \Fig{Fig:XCChiral}, with a bond dimension up to $D^\ast=8192$ SU(2) 
multiplets 
(corresponding to $D> 22\, 000$ 
individual states), 
{chiral correlations are strongly suppressed.
{We further perform a linear $1/D^\ast\to0$ extrapolation from 
the correlation data of the largest three bond dimensions $D^\ast=4096,5793,8192$, 
and still see no sign of long-range chiral correlation.}
Hence XC4 behaves very differently from YC4 or YC6
where for cylinders of the same length already robust
long-range chiral correlations were observed, having
$\chi^2 =  0.128$ for YC4 at $D^\ast=4096$ [\Fig{Fig:Chiral}],
or $\chi^2 \approx 0.25$ for YC6 at $D^\ast \gtrsim 8192$
[\Fig{fig:YC6:cmplx}].
Here for XC4, the chiral correlations drop rapidly over
short distances $d_{ij}<10$. However, they appear to gain
weak support for $d_{ij}>10$ around the much smaller value
$\chi^2 \sim 10^{-4}$. While the long-distance
correlations still gain strength with increasing \Dstar,
nevertheless,
the chiral correlations do show (weak) decay with distance,
even within the numerically converged range
$d_{ij}\lesssim 25$. In this sense, we see no clear support for}
long-range chiral correlation in XC4 cylinders.}

\end{appendices}

\bibliography{./lib/TLU.bib}

\end{document}